\documentclass[12pt]{article}
\usepackage{hyperref}
\usepackage{amsmath,amssymb}
\usepackage{url}
\usepackage{graphicx,color}
\usepackage{multirow,array}
\newcommand{\Exp}[1]{\exp{\biggl(#1 \biggr)}}
\newcommand{\PA}[1]{\biggl(#1 \biggr)}
\newcommand{\bb}[1]{\mathbf{#1}}
\newcommand{\bbb}[1]{\mathbb{#1}}

 \def\ep{{\epsilon}}

 \def\frac#1#2{{#1\over #2}}

 \def\s{\sqrt}

\def\be{\begin{equation}}
\def\ee{\end{equation}}
\def\ba{\begin{eqnarray}}
\def\ea{\end{eqnarray}}
\numberwithin{equation}{section}

 \def\de{\partial}

 \def\ti{\tilde}
 
  \def\al{\alpha'}

 \def\ddd{\cdot\cdot\cdot}
 \def\no{\nonumber \\}

 \def\la{\langle}
 \def\lb{\rangle}
 \def\ep{\epsilon}

\begin{document}

\begin{titlepage}
\thispagestyle{empty}

\begin{flushright}
NORDITA-2017-1
\\
YITP-17-1
\\
IPMU17-0003
\\
\end{flushright}

\bigskip

\begin{center}
\noindent{{ \textbf{Evolution of Entanglement Entropy in Orbifold CFTs}}}\\
\vspace{2cm}
Pawel Caputa$^{a,b}$, Yuya Kusuki$^{b}$, Tadashi Takayanagi$^{b,c}$ and Kento Watanabe$^{b}$
\vspace{1cm}

{\it
$^{a}$ Nordita, KTH Royal Institute of Technology and Stockholm University,
Roslagstullsbacken 23, SE-106 91 Stockholm, Sweden\\
$^{b}$Center for Gravitational Physics, Yukawa Institute for Theoretical Physics,\\
Kyoto University, 
Kyoto 606-8502, Japan\\
$^{c}$Kavli Institute for the Physics and Mathematics of the Universe,\\
University of Tokyo, Kashiwa, Chiba 277-8582, Japan\\
}

\vskip 2em
\end{center}

\begin{center}
{\it Dedicated to John Cardy on his 70th birthday}
\end{center}

\begin{abstract}
In this work we study the time evolution of Renyi entanglement entropy for locally excited states created by twist operators in cyclic orbifold $(T^2)^n/\bbb{Z}_n$ and symmetric orbifold $(T^2)^n/S_n$. We find that when the square of its compactification radius is rational, the second Renyi entropy approaches a universal constant equal to the logarithm of the quantum dimension of the twist operator. On the other hand, in the non-rational case, we find a new scaling law for the Renyi entropies given by the double logarithm of time $\log\log t$ for the cyclic orbifold CFT.
\end{abstract}

\end{titlepage}

\newpage

\section{Introduction}

The most fundamental quantity which characterizes the degrees of freedom in conformal field theories (CFTs) is the thermal entropy, which can be computed universally in two dimension thanks to the celebrated Cardy formula \cite{Cardy:1986ie}. Entanglement entropy (EE) provides a more general probe of CFTs \cite{HLW, Cardy} and can even capture various dynamical processes in CFTs. Especially, as pioneered by Calabrese and Cardy, quantum quenches provide very important classes of excited states in CFTs and beautiful results on the time evolution of entanglement entropy have been derived \cite{GQ,CCL,CCR,CCR2} for two dimensional CFTs. A quantum quench is triggered by a sudden shift of Hamiltonian at a specific time. This shift can happen either globally or locally, and we talk about global quench or local quench, respectively. To analyze quantum quenches in higher dimensional CFTs, holographic entanglement entropy \cite{RT1,RT2,RT3,RTR1,RTR2,RTR3} also provides a useful tool \cite{AAL,Ba,HaMa}.

There is another interesting class of excited states in CFTs which are simple enough to be computed analytically. These are locally excited states obtained by acting a local operator $O(x)$ on the vacuum in a given CFT at the time $t=0$, introduced in \cite{NNT} for the purpose of computation of entanglement entropy. The state at time $t$ is explicitly written as
\be
|\psi\lb={\cal N}\cdot e^{-i t  H} \cdot e^{-\ep H}\cdot O(x_*)|0\lb,  \label{localex}
\ee
where $x_*$ represents the position of insertion of the operator, $\ep$ is an UV regularization of the local operator and ${\cal N}$ is a normalization factor so that $\la\psi|\psi\lb=1$.

By applying the replica method, we can calculate the time evolution of entanglement entropy
and its generalization called Renyi entanglement entropy (Renyi EE). To define these quantities we trace out a subsystem $B$, and define a reduced density matrix $\rho_A$ for the subsystem $A$, which is the complement of $B$. In this work, we will simply set $A$ to be a half space.
The $m$-th Renyi entropy $S^{(m)}_A$ is defined by
\ba
S^{(m)}_A=-\frac{1}{m-1}\log \mbox{Tr}[\rho_A^m].
\ea
The limit $m\to 1$  defines the (von-Neumann) entanglement entropy $S_A$.
For our excited state (\ref{localex}), the computation of $S^{(m)}_A$ is equivalent to
that of the $2m$-point function on the ($m$ times) replicated space \cite{NNT,Nozaki2014}.
Our main focus will be on the difference $\Delta S^{(m)}_A$ between the entropy for a given excited state and the vacuum state so that the area law UV divergences are cancelled.

In previous works, several interesting features of $\Delta S^{(m)}_A$ have been worked out for excited states in CFTs of the form (\ref{localex}) and our primary interest in this paper is to proceed more in this direction for two dimensional CFTs. First, it was found that the Renyi EE growth $\Delta S^{(m)}_A$ approaches to a finite value at late time in free CFTs in any dimensions \cite{NNT,Nozaki2014,Nozaki:2015mca,Caputa2016b,Nozaki:2016mcy} and in (two dimensional) rational CFTs (RCFTs) \cite{HNTW,CV,CGHW,Caputa2016yzn,Numasawa2016}. On the other hand, the holographic result \cite{NNTH,CNT} for two dimensional CFTs shows the logarithmic growth $\Delta S^{(m)}_A\sim \frac{c}{6}\log t$ under the time evolution. This holographic behavior was reproduced from a CFT computation by utilizing the spectrum gap and the known behavior of conformal blocks and in the large central charge limit \cite{ABGH}  (see also \cite{Caputa2015b,Caputa2015c} for further generalizations to finite temperature).

For two dimensional CFTs, these previous results cover two extremal cases: rational CFTs and holographic CFTs. Note that the latter CFTs are expected both to be strongly coupled and to have large central charges so that they are dual to classical gravity on AdS$_3$. One may naively think that the behavior of such time evolutions depend on whether the CFT is integrable or chaotic. This speculation raises a question: do we only have two possible behaviors of $\Delta S^{(m)}_A$ for any two dimensional CFTs, i.e. (i) approaching a finite constant (as in RCFTs) and (ii) growing logarithmically (as in holographic CFTs) ?
This question motivates us to study integrable CFTs which are not rational. For this purpose we would like to study $\Delta S^{(m)}_A$ for a class of solvable CFTs defined by the sigma model whose target space is the cyclic orbifold:
\be
(T^2)^n/\bbb{Z}_n,  \label{sycft}
\ee
where $T^2=S^1\times S^1$ is the $c=2$ CFT defined by two compact bosons $X_1$ and $X_2$, both of which are compactified on the same radius $R$. The $\bbb{Z}_n$ action is defined by shifting $n$ copies of the two dimensional torus $T^2$ successively. We will choose the primary operator $O$ in (\ref{localex}) which creates the local excitation, to be the twist operator $\sigma_n$. Thanks to analytical results by Calabrese, Cardy and Tonni \cite{CCT}, we can have an analytical expression for $\Delta S^{(m)}_A$ in this CFT model.

When $R^2$ is rational, the $c=2$ CFT becomes a rational CFT. Moreover, as we will see later, the cyclic orbifold CFT (\ref{sycft}) also becomes rational for any $n$.  However, if $R^2$ is irrational, these CFTs are irrational. Therefore this offers an example of integrable but irrational CFTs. As we will show in this paper, the 2nd Renyi EE $\Delta S^{(2)}_A$  in this irrational CFT actually shows a new behavior under the time evolution different from (i) and (ii).

A similar statement is true for the symmetric orbifold CFT defined by a sigma model whose target space is given by
\be
(T^2)^n/S_n,  \label{symcft}
\ee
where $S_n$ is the symmetric group. When the square of compactification radius $R^2$ is rational (or irrational), this symmetric orbifold CFT is also rational (irrational). Note that only when $n=2$, (\ref{sycft}) and (\ref{symcft}) are equivalent. This class of CFTs is also motivated by the fact that typical examples of AdS$_3/$CFT$_2$ are given by symmetric orbifold CFTs of the form $M^n/S_n$ for various choices of 2d CFTs $M$, though we need to deform them by exactly marginal operators to reach a CFT which has a classical gravity dual. Refer to \cite{Balasubramanian:2016xho} for an interesting generalization of entanglement entropy (called entwinement) of the ground state in a symmetric orbifold CFT and its connection to AdS$_3/$CFT$_2$. In this paper, we will indirectly compute $\Delta S^{(m)}_A$ for the excited state created by the twist operator in the rational case using the connection to the quantum dimensions \cite{HNTW}, which in general leads to different results than those for the cyclic orbifold.

This paper is organized as follows:
In section two, we review the calculation of (2nd) Renyi entanglement entropy in two dimensional CFTs and useful results in our symmetric orbifold CFTs.
In section three, we present our results for Renyi EE when the cyclic orbifold CFT becomes rational and irrational respectively. In section four, we discuss the computation of Renyi EE in the symmetric orbifold CFT. In section five, we re-interpret the results in terms of mutual information for light-like separated intervals.
In section six, we summarize our conclusions.
In the appendix A, we present a summary of the computation of quantum dimensions.
In the appendix B, we explain the details of the computation of a determinant.

\section{$\Delta S^{(2)}_A$ and four point functions}

We begin by reviewing the computations of the second Renyi entropy $\Delta S^{(2)}_A$ \cite{HNTW} and four point functions \cite{CCT} in the cyclic orbifold CFTs in a way convenient for our later analysis.

\subsection{2nd Renyi EE from 4-pt function}

\begin{figure}[htbp]
  \centering
  \includegraphics[width=8cm]{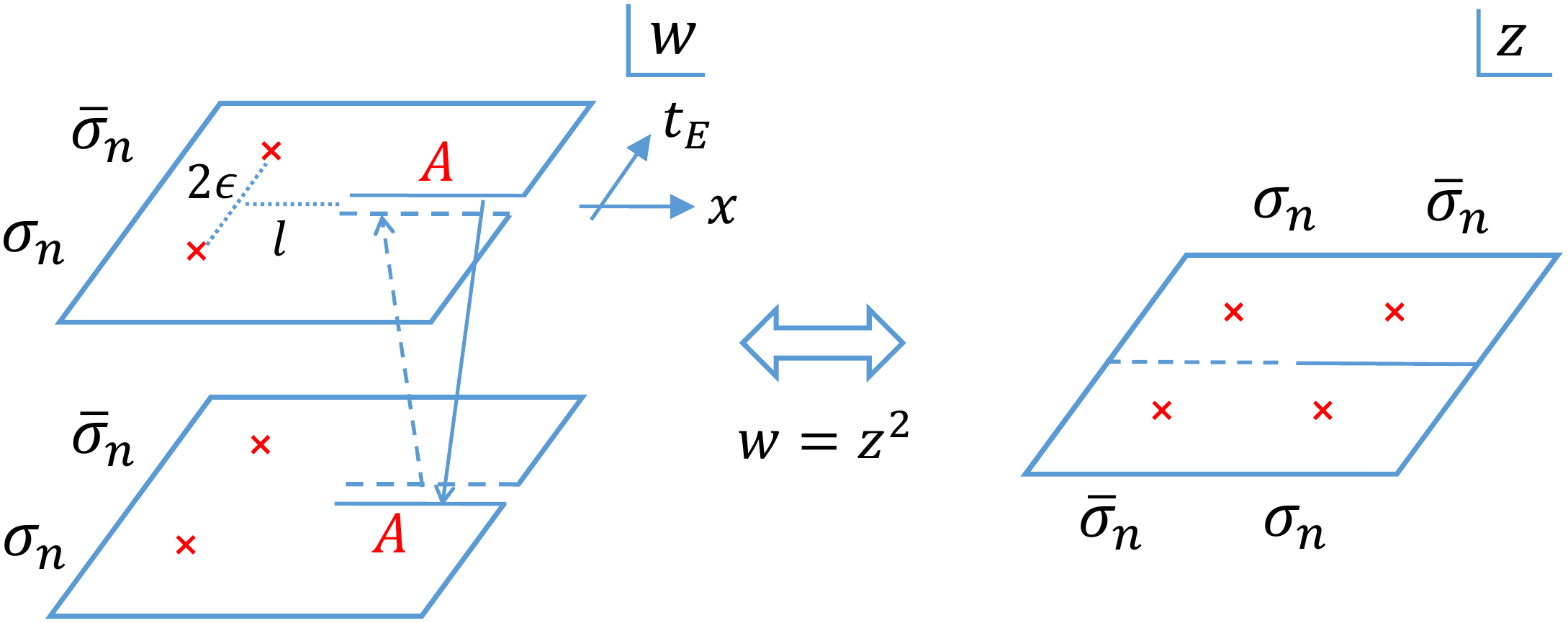}
  \caption{The setup of the 2nd Renyi entanglement entropy $\Delta S^{(2)}_A$ for a half line $A$}\label{Setupfig}
\end{figure}

 We would like to compute the growth of the 2nd Renyi entanglement entropy
$\Delta S^{(2)}_A$ for an excited state of the form (\ref{localex}).
We describe the 2d Euclidean space $\bbb{R}^2$ by a complex coordinate $(w,\bar{w})=(x+it_E,x-it_E)$, where $t_E$ is the Euclidean time and $x$ is the space coordinate. We choose the location of operator insertion to be $x_*=-l<0$ at $t=0$ and to be smeared by $\epsilon$.
We take the subsystem $A$ to be the half of the space $x>0$.
To calcualte $\Delta S^{(2)}_A$, we replicate the $w$-plane into 2-sheets, glue them along $A$
and uniformize it by the conformal map $w = z^2$ as in Fig.\ref{Setupfig}.
As showed in \cite{HNTW}, we can calculate $\Delta S^{(2)}_A$ for the excited state (\ref{localex}) from the four point function of
the operator $O$. The relevant four point function in our CFT has the following structure:
\ba
\la O(z_1,\bar{z}_1)\bar{O}(z_2,\bar{z}_2)O(z_3,\bar{z}_3)\bar{O}(z_4,\bar{z}_4)\lb
=|z_{13}z_{24}|^{-4\Delta}|z|^{-4\Delta}|1-z|^{-4\Delta}F_O(z,\bar{z}),
\ea
where $z$ is the cross ratio $z=\frac{z_{12}z_{34}}{z_{13}z_{24}}$.

The growth of 2nd Renyi entanglement entropy for $t>l$ is given by
\be
\Delta S^{(2)}_A=-\lim_{(z,\bar{z})\to (1,0)}\log F_O(z,\bar{z}),
\ee
where the limit is understood more precisely as follows: for an infinitesimally small $\ep$, we have (assuming $t\gg l$)
\be
z\simeq 1-\frac{\ep^2}{4t^2},\ \ \  \bar{z}\simeq \frac{\ep^2}{4t^2}. \label{limep}
\ee
In 2d rational CFTs, we can take $\ep\to 0$ limit directly and we end up with
\be
\label{qdim}
\Delta S^{(2)}_A=\log d_{O},
\ee
where $d_{O}$ is the quantum dimension of the primary operator $O$ \cite{HNTW}.
It is defined by the ratio of the elements of the modular S-matrix
\be
d_O=\frac{S_{0O}}{S_{00}}, \label{do}
\ee
where $0$ denotes the identity sector. Moreover, a more general analysis shows that for any $m$,
the $m$-th Renyi entropy growth $\Delta S^{(m)}_A$ also takes the same value 
\be
\Delta S^{(m)}_A=\log d_{O}, \label{qqdim}
\ee
as proved in \cite{HNTW}.

In this paper we focus on the cyclic orbifold CFTs given by (\ref{sycft}) with the central charge $c=2n$ and we choose the primary operator $O$ to be the twist operator 
\be
O(z,\bar{z})=\sigma_n(z,\bar{z}),
\ee
where $\sigma_n$ is the twisted operator for the cyclic transformation
$X^{(i)}_{1,2}\to X^{(i+1)}_{1,2}$ for $i=1,2,\ddd,n$. 
The conformal dimension of $\sigma_n$ is given by $\Delta_n=\frac{1}{12}(n-1/n)$.

Its four point function takes the form
\footnote{In a recent work \cite{Shiba:2017vsr}, similar four point functions of twist operators were considered for the analysis of the Aharonov-Bohm effect on entanglement entropy in 2d CFT.}:
\ba
\la \sigma_n(z_1,\bar{z}_1)\bar{\sigma}_n(z_2,\bar{z}_2)\sigma_n(z_3,\bar{z}_3)
\bar{\sigma}_n(z_4,\bar{z}_4)\lb
=|z_{13}z_{24}|^{-4\Delta_n}|z|^{-4\Delta_n}|1-z|^{-4\Delta_n}F_n(z,\bar{z}), \label{fnd}
\ea
where $z$ is the cross ratio $z=\frac{z_{12}z_{34}}{z_{13}z_{24}}$. This way, in the states excited by twist operators, the growth of 2nd Renyi EE for $t>l$ is found from
\be
\Delta S^{(2)}_A=-\lim_{(z,\bar{z})\to (1,0)}\log F_n(z,\bar{z}).
\ee
To fix our conventions, we denote the radius of $T^2$ as $R$. As in \cite{CCT}, we introduce the parameter $\eta=R^2$, where we chose the action of the free scalar which describes $T^2$ as\footnote{We set $g=1$ in the notation of
\cite{CCT}. In other words, our convention corresponds to $\al=1$ in string theory.}
\be
S=\frac{1}{4\pi}\int dtdx (\de_\mu\phi)^2.
\ee
Our convention is then such that $\eta=1$ corresponds to the self-dual radius and $\eta=1/2$ is equivalent to a Dirac fermion.

\subsection{Four Point Functions in $(T^2)^n/\bbb{Z}_n$}

Here we summarize the known expression for the four point function in the cyclic orbifold CFT $(T^2)^n/\bbb{Z}_n$ following \cite{CCT}. 
As in (\ref{fnd}), the correlator is characterized by the function
$F_n(z,\bar{z})$ which is expressed as follows \cite{CCT}:
\ba
F_n(z,\bar{z})\equiv \frac{2^{n-1}\eta^{n-1}}{\prod_{k=1}^{n-1}I_{k/n}(z,\bar{z})}\cdot \Theta\left(0|\eta\Gamma\right)^2, \label{xxq}
\ea
where the $2(n-1)\times 2(n-1)$ symmetric matrix $\Gamma$ is defined by
\be
\Gamma=\left(
  \begin{array}{cc}
    i\Omega & -\Lambda/2 \\
    -\Lambda^T/2 & i\ti{\Omega} \\
  \end{array}
\right),
\ee
and the $p$ dimensional theta function is defined as
\be
\Theta(0|\Gamma)=\sum_{m\in \bbb{Z}^p}e^{i\pi m^T\cdot\Gamma\cdot m}.
\ee
We also introduced
\ba
\label{eq:Ikn}
I_{k/n}(z,\bar{z})=f_{k/n}(z)\bar{f}_{k/n}(1-\bar{z})+\bar{f}_{k/n}(\bar{z})
f_{k/n}(1-z),
\ea
where
\be
f_{k/n}(z)=\,_2F_1(k/n,1-k/n,1,z).
\ee
Note the identities $f_{k/n}(0)=1$ and $f_{0}(z)=1$.\\
The $(n-1)\times (n-1)$ matrices $\Lambda$ and $\Omega$, $\ti{\Omega}$ are defined by
\ba
\label{eq:Mat}
&& \Lambda_{r,s}=\frac{4}{n}\cdot \sum_{k=0}^{n-1}
b_k\sin\left(\frac{\pi k}{n}\right)\cdot \sin\left(\frac{2\pi}{n}k(r-s+1/2)\right),\no
&& \Omega_{r,s}=\frac{2}{n}\cdot \sum_{k=0}^{n-1}a_k
\sin\left(\frac{\pi k}{n}\right)\cdot \cos\left(2\pi\frac{k}{n}(r-s)\right),\no
&& \tilde{\Omega}_{r,s}=\frac{2}{n}\cdot \sum_{k=0}^{n-1}\ti{a}_k
\sin\left(\frac{\pi k}{n}\right)\cdot \cos\left(2\pi\frac{k}{n}(r-s)\right),
\ea
where $r, s = 1, \cdots , n-1$ and
\ba
\label{eq:mat}
&& a_k=\frac{2f_{k/n}(1-z)\bar{f}_{k/n}(1-\bar{z})}{I_{k/n}(z,\bar{z})}, \no
&& \ti{a}_k=\frac{2f_{k/n}(z)\bar{f}_{k/n}(\bar{z})}{I_{k/n}(z,\bar{z})}, \no
&& b_k=\frac{f_{k/n}(z)\bar{f}_{k/n}
(1-\bar{z})-f_{k/n}(1-z)\bar{f}_{k/n}(\bar{z})}{I_{k/n}(z,\bar{z})}.
\ea
The overall normalization of (\ref{xxq}) is chosen such that
\be
F_n(0,0)=1.
\ee
Also the four point function $F_n(z,\bar{z})$ manifestly satisfies the channel duality relation
\be
F_n(z,\bar{z})=F_n(1-z,1-\bar{z}),
\ee
which is the invariance under the transformation $z \leftrightarrow 1-z, \bar{z} \leftrightarrow 1-\bar{z}$.
Under this transformation, the theta function $\Theta\left(0|\eta\Gamma\right)$ is itself invariant.

\subsection{Another Expression}

It is possible to write the four point function $F_n(z,\bar{z})$
in a manifestly invariant form under $\eta \leftrightarrow 1/\eta$.
By performing the Poisson resummation,
we can derive the transformation of
the theta function $\Theta\left(0|\eta\Gamma\right)$
under $\eta \leftrightarrow 1/\eta$
\cite{Calabrese2013,Coser2014}.
\be
\Theta(0|\eta\Gamma) = \eta^{-(n-1)}\cdot \Theta(0|\Gamma/\eta) .
\ee
Applying this formula to (\ref{xxq}),
we can obtain the following interesting relation
\ba
\frac{F_n(z,\bar{z})}{F^{(\eta=1)}_n(z,\bar{z})}=
\frac{ \Theta (0 | \eta \Gamma) \cdot \Theta (0 | \Gamma/\eta)}{ \left( \Theta (0 | \Gamma) \right)^2},  \label{idef}
\ea
where $F^{(\eta=1)}_n(z,\bar{z})$ is the function $F_n(z,\bar{z})$ at the self-dual point $\eta=1$.
\ba
F^{(\eta=1)}_n(z,\bar{z}) =
\frac{2^{n-1}}{\prod_{k=1}^{n-1}I_{k/n}(z,\bar{z})}\cdot \Theta\left(0|\Gamma\right)^2.
\ea
Note also the special property that when $z=\bar{z}=x$ ($x$ is a real number) we simply find
\be
F^{(\eta=1)}_n(x,x)=1.
\ee
The expression (\ref{idef}) manifestly shows that
$F_n(z,\bar{z})$ is invariant under the T-duality $\eta\leftrightarrow 1/\eta$
and the $\eta$ dependence of $F_n(z,\bar{z})$ comes from the ratio (\ref{idef}).

\subsection{Generalization to Two Different Radii}

When the two radii of $T^2$ are different, we can generalize our previous result as follows
\begin{align}
F_n (z, \bar{z})
&=
\frac{2^{n-1}}{\prod_{k=1}^{n-1}I_{k/n}(z,\bar{z})}\cdot
\prod_{q=1,2} \left[ \eta_q^{\frac{n-1}{2}} \cdot \Theta \left(0 | \eta_q \Gamma \right)  \right],
\end{align}
where $\eta_q = R_q^2 \ (q=1, 2)$.
The overall normalization is fixed such that $F_n (0, 0) = 1$  as the previous case ($\eta = \eta_1=\eta_2$).\\
The four point function manifestly satisfies
the invariance under  $z \leftrightarrow 1-z, \bar{z} \leftrightarrow 1-\bar{z}$
and also under $\eta_1 \leftrightarrow \eta_2$.
By using the Poisson resummation formula, we can confirm the invariance under $\eta_q  \leftrightarrow 1/\eta_q$.
Our discussions in the following sections can be applied straightforwardly to the two different radii case.

\section{Growth of Renyi Entanglement Entropy}

In this section we will turn to the computation of $\Delta S^{(2)}_A$ in $(T^2)^n/\bbb{Z}_n$. We start by the analytical computation of $F_n(1-\epsilon^2/4 t^2,\epsilon^2/4 t^2)$ in the limit $\epsilon \to 0$. We then find that the behavior of $F_n$ (i.e. $\Delta S^{(2)}_A$) for irrational $\eta$ is distinctly different from that for rational $\eta$.

For rational $\eta \  ( = \frac{p}{q} )$, the CFT is rational and, as expected from the result in \cite{HNTW}, $\Delta S^{(2)}_A$ approach to a finite constant that we prove to be
\be
\Delta S^{(2)}_A=(n-1)\cdot\log (2pq).
\ee
We derive this result by both, analytical and numerical computations
and ensure consistency with the formula (\ref{qdim}) by using the quantum dimension of the twist operator $\sigma_n$ evaluated in the Appendix \ref{kusuki}.

For irrational $\eta$, we encounter a new late time behavior of  $\Delta S^{(2)}_A$ in the form of the double logarithm
\be
\Delta S^{(2)}_A\simeq (n-1)\cdot \log\left(\log(t/\ep)\right).
\ee
This evolution belongs to neither the RCFT class nor the holographic CFT class encountered before and is the main new result of our work. This late time scaling suggest the existence of the third class, {\it irrational CFT class} from the perspective of the evolution of entanglement measures in excited states.

\subsection{Analytical Computation of $F_n(1,0)$}

Now let us closely study the $\ep\to 0$ limit (\ref{limep}). It is useful to
define an infinitesimal quantity $\delta$
\be
\delta\equiv \frac{\pi}{\log (4t^2/\ep^2)} \to 0.
\ee
In this limit (\ref{limep}) with $\delta \to 0$,
(\ref{eq:Ikn}) and (\ref{eq:mat}) are approximated by
\ba
&&I_{k/n}\simeq \frac{\sin^2\left(\frac{\pi k}{n}\right)}{\delta^2}+O(\delta^0),\no
&&a_k\simeq\frac{2\delta}{\sin\left(\frac{\pi k}{n}\right)}+O(\delta^2),\no
&&\ti{a}_k\simeq\frac{2\delta}{\sin\left(\frac{\pi k}{n}\right)}+O(\delta^2),\no
&&b_k\simeq1+O(\delta^2),
\ea
and (\ref{eq:Mat}) are also approximated by
\ba
\label{eq:NMat}
&&\Omega_{r,s}\simeq \frac{4}{n}\delta\cdot (-1+n\cdot\delta_{r,s}) \ (\equiv \delta \cdot(\Omega_0)_{r,s}), \no
&&\ti{\Omega}_{r,s}\simeq \frac{4}{n}\delta\cdot (-1+n\cdot\delta_{r,s})\ (\equiv \delta \cdot(\Omega_0)_{r,s}), \no
&&\Lambda_{r,s}\simeq2(\delta_{r,s}-\delta_{r,s-1})\ (\equiv (\Lambda_0)_{r,s}).\no
\ea
Therefore we can get the simple approximation form of the function $F_n$ (\ref{xxq}) as
\begin{equation}
F_n\simeq 2^{n-1}\eta^{n-1}\cdot (g_n)^2,
\end{equation}
with
\begin{equation}
g_n\equiv \frac{2^{n-1}\delta^{n-1}}{n}\cdot \sum_{\bb{l},\bb{m} \in {\bbb{Z}}^{n-1}}
e^{- \pi \delta \eta \bb{m}^T \cdot \Omega_0 \cdot \bb{m}  + 2 \pi i \eta (-\frac{1}{2}\bb{l}^T\cdot\Lambda_0^T)\cdot \bb{m} - \pi \delta\eta \bb{l}^T \cdot (\Omega_0) \cdot \bb{l} },
\end{equation}
where we useed the following relation
\begin{equation}
\prod_{k=1}^{n-1}2\sin\biggl(\frac{\pi k}{n}\biggr)=n.
\end{equation}
By using the Poisson resummation formula
\ba
\sum_{\bb{m} \in {\bbb{Z}^{n-1}}}
e^{- \pi \bb{m}^T\cdot A \cdot \bb{m} + 2 \pi i \bb{b}^T \cdot \bb{m} }
=
\frac{1}{\sqrt{\det A}} \sum_{\bb{\ti{m}} \in {\bbb{Z}^{n-1}}}
e^{- \pi (\bb{\ti{m}} + \bb{b} )^T \cdot (A^{-1}) \cdot (\bb{\ti{m}} + \bb{b} ) },
\ea
we find
\footnote{Here we omit the $O(\delta)$ subleading corrections to the quadratic form on the exponent.
They don't play an important role in the following discussion.}
\begin{align}
\label{eq:gn}
g_n &=
\frac{2^{n-1}\delta^{n-1}}{n} \cdot
\frac{1}{\sqrt{\det ( \eta \delta \Omega_0 ) } }
\sum_{\bb{l},\bb{\ti{m}} \in {\bbb{Z}^{n-1}}}
e^{- \frac{\pi}{\eta \delta} (\bb{\ti{m}}^T - \frac{\eta}{2} \bb{l}^T \cdot\Lambda_0^T ) \cdot \Omega_0^{-1} \cdot (\bb{\ti{m}} -\frac{\eta}{2} \Lambda_0 \cdot \bb{l} ) -\pi \eta \delta \bb{l}^{T} \cdot \Omega_0 \cdot  \bb{l}}.
\end{align}
Since $\Omega_0^{-1}$ is Hermitian it may be diagonalized by a unitary matrix $U$: $\Omega_0^{-1}=U D U^{\dagger} $. If $\bb{x}=U\bb{y}$, then
\begin{equation}
\bb{x}^T \cdot \Omega_0^{-1} \cdot \bb{x}=\sum_i D_{ii} |y_i|^2,
\end{equation}
hence if $D_{ii}>0$, then $\bb{x}^T \cdot \Omega_0^{-1} \cdot \bb{x}>0$ for all $\bb{x},\bb{x}\neq\bb{0}$ and we can easily show $D_{ii}>0$ (see Appendix \ref{kusuki2}). This discussion tells us that
the quadratic form
$ (\bb{\ti{m}}^T - \frac{\eta}{2} \bb{l}^T \cdot\Lambda_0^T ) \cdot \Omega_{\delta}^{-1} \cdot (\bb{\ti{m}} -\frac{\eta}{2} \Lambda_0 \cdot \bb{l} ) $
in the exponent  is minimized by the following condition
\ba
\label{eq:0con}
\bb{\ti{m}} -\frac{\eta}{2} \Lambda_0 \cdot \bb{l} = \bb{0}.
\ea
This gives us the dominant contribution in the $\delta \to 0$ limit.
For rational $\eta$($= \frac{p}{q}$),
the condition is satisfied by
\begin{equation}
\ti{m}_r=pk_r
\end{equation}
and
\begin{equation}
\begin{aligned}
\PA{\frac{1}{2}\Lambda_0\cdot l}_r=l_r-l_{r+1}&=qk_r \ \ \ \ (r=2,\ldots,n-2)\\
-l_1&=qk_1\\
l_{n-1}&=qk_{n-1},
\end{aligned}
\end{equation}
where $k_r$ is an arbitrary integer. Then, we can estimate $g_n$ for rational $\eta$ by applying this condition
\begin{equation}
\begin{aligned}
g_n  &=
\frac{2^{n-1}\delta^{n-1}}{n} \cdot
\frac{1}{\sqrt{\det ( \eta \delta \Omega_0 ) } }
\sum_{\bb{k} \in {\bbb{Z}}^{n-1}}
 e^{-\pi p q \delta \bb{k}^{T} \cdot \Omega_0 \cdot  \bb{k}}\\
&=
\frac{2^{n-1}\delta^{n-1}}{n} \cdot
\frac{1}{\sqrt{\det ( \eta \delta \Omega_0 \cdot p q \delta \Omega_0 ) } } \\
&=
\frac{2^{n-1}}{n p^{n-1} \det{\Omega_0}} \\
&=
\frac{1}{(2p)^{n-1}} .
\end{aligned}
\end{equation}
where we used the following relation
\begin{equation}
\label{eq:Omega}
\det \Omega_0=\frac{2^{2(n-1)}}{n}.
\end{equation}
This relation is derived in Appendix \ref{kusuki2}.
Finally, we can obtain the four point function $F_n$ for rational $\eta (= \frac{p}{q})$.
\begin{align}
\label{eq:rational}
F_n
\simeq 2^{n-1} \eta^{n-1} \cdot \frac{1}{(2 p)^{2(n-1)}}
= \frac{1}{(2 p q)^{n-1}} .
\end{align}
On the other hand, for irrational $\eta$,
only $\tilde{m}_r = {l}_r = 0$ satisfies the condition (\ref{eq:0con}). Therefore we have
\begin{equation}
\begin{aligned}
g_n &=
\frac{2^{n-1}\delta^{n-1}}{n} \cdot
\frac{1}{\sqrt{\det ( \eta \delta \Omega_0 ) } } \\
 &=
\sqrt{\frac{\delta^{n-1}}{n \eta^{n-1}}},
\end{aligned}
\end{equation}
and we approximate $F_n$ for irrational $\eta$ as
\begin{equation}
\label{eq:irrational}
F_n
\simeq 2^{n-1} \eta^{n-1} \cdot \frac{\delta^{n-1}}{n \eta^{n-1}}
= \frac{2^{n-1}}{n} \delta^{n-1}.
\end{equation}
It is very important to stress that this approximation is only applicable only when
\be
\delta \ll\frac{1}{pq}.
\ee
If we for example consider the case (assuming $q\gg1$)
\be
\frac{1}{pq}\ll \delta \ll 1,
\ee
we cannot justify the approximation that only $(\ti{m}_r,l_r-l_{r+1})=(p,q)k_r$ contributes.

\subsection{Growth of Renyi Entanglement Entropy for rational $\eta$}

In the rational case the four point function $F_n(1,0)$ is written by (\ref{eq:rational}), hence the 2nd Renyi entropy increases by the constant equal to
\be
\label{eq:Renyi}
\Delta S^{(2)}_A=(n-1)\cdot\log (2pq),
\ee
which can be as large as the central charge $c=2n$ in the large $n$ limit.

\begin{figure}
  \centering
  \includegraphics[width=5cm]{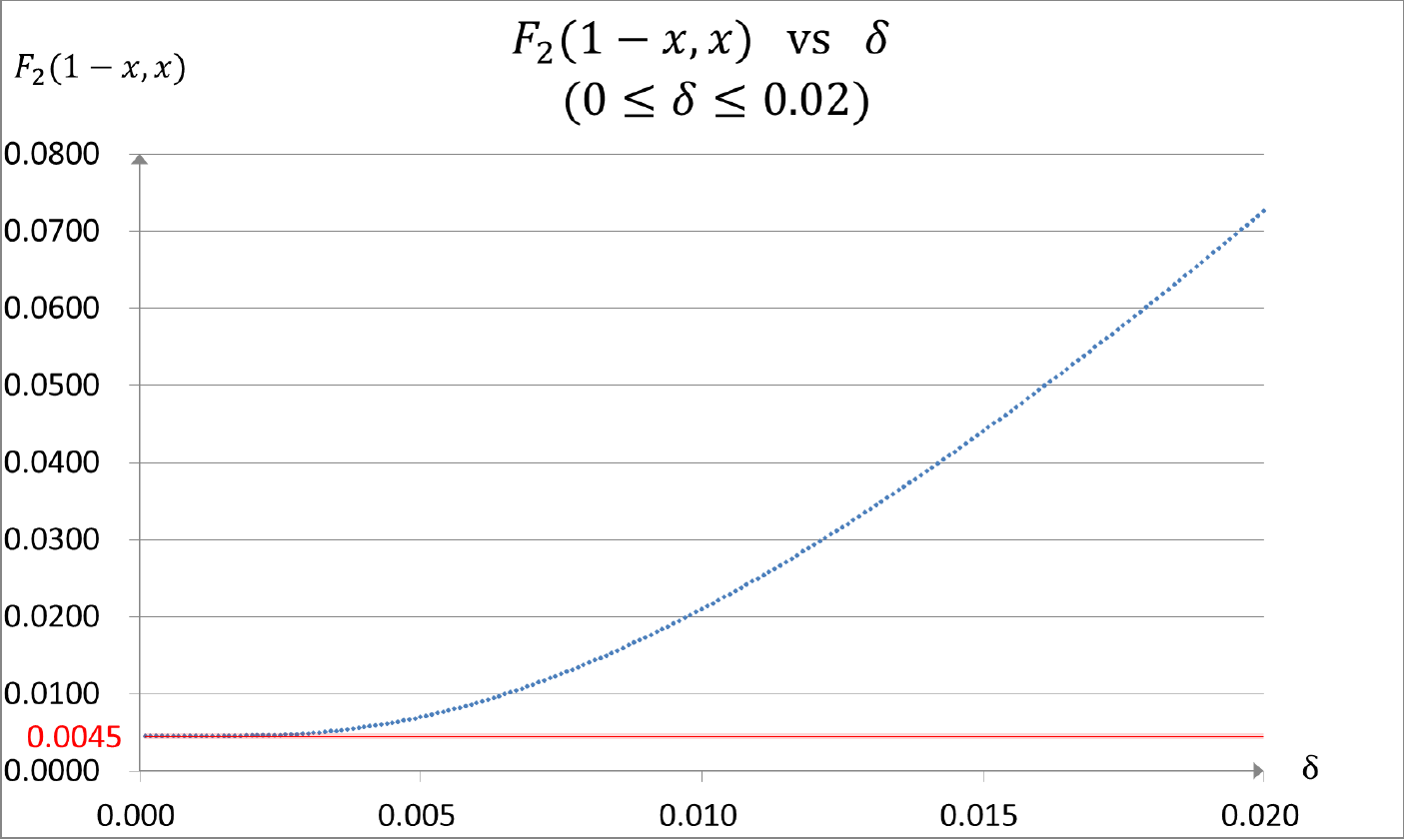}
  \includegraphics[width=5cm]{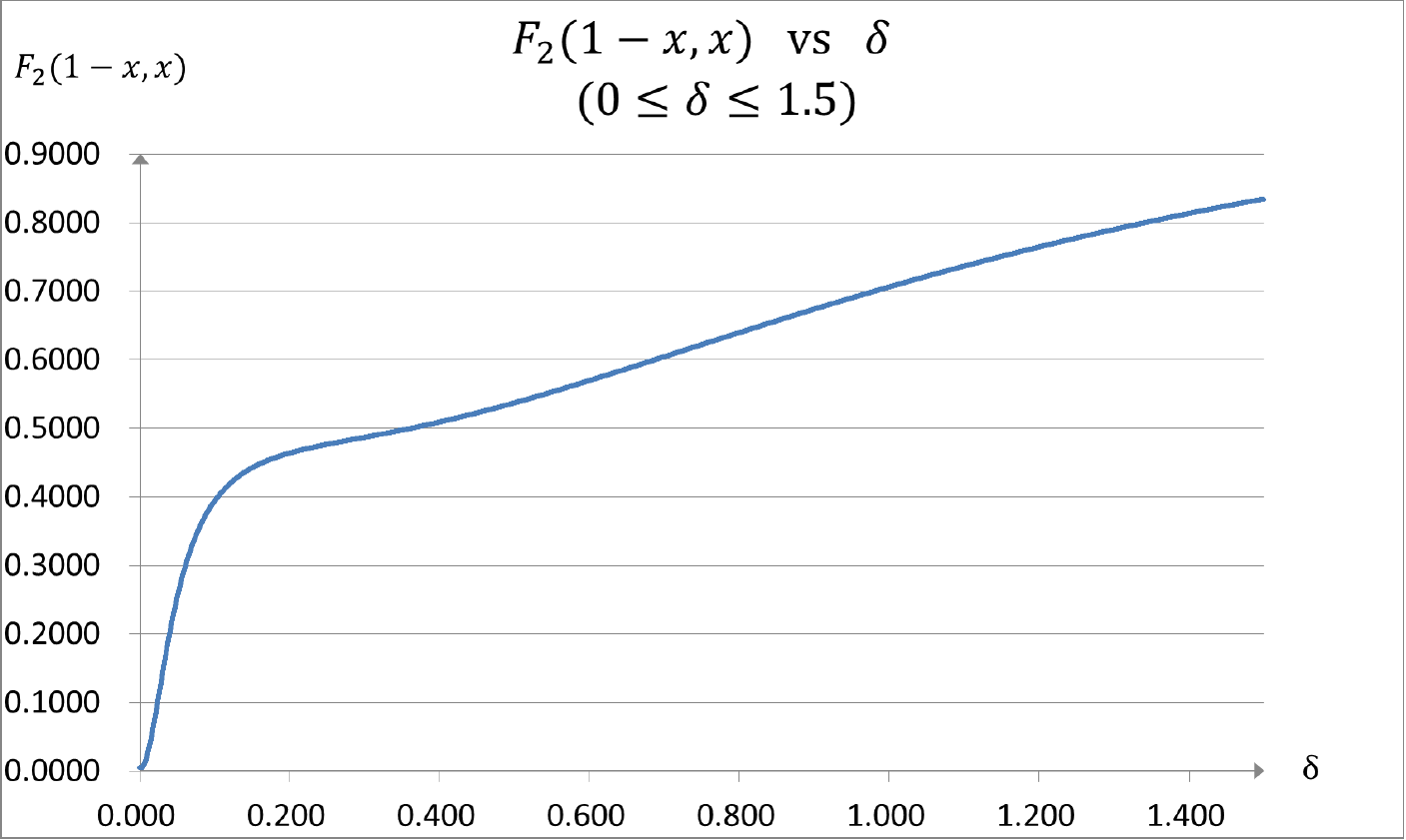}
  \caption{The plot of the function $F_2(x,1-x)$ for $\eta=\frac{10}{11}$ as a function of $\delta=\frac{\pi}{-\log x}$. The left figure for $0\leq \delta\leq 0.02$. The right one for $0\leq \delta\leq 1.5$. }\label{1011fig}
  \end{figure}

\medskip
A few comments to support this result are in order at this point :
\begin{itemize}
    \setlength{\leftskip}{0cm}
    \item First of all, we can check the formula (\ref{eq:Renyi}) numerically for various values of $(z,\bar{z})$ and $(p,q)$. For instance, the figure \ref{1011fig} shows the behavior of $F_2$ for
$\eta=\frac{10}{11}$ as a function of $\delta$. Indeed $F_2$ approaches to $\frac{1}{220}$ in the $\delta \to 0$ limit.  Also there is a plateaux $F_2\simeq 1/2$ for $\delta \sim 0.3$ as $\eta$ is close to $1$. A similar plateaux can be observed more clearly for $\eta=\frac{10^p}{10^p+1}$ for a large $p$.\\
    \item When $\eta$ is rational, we expect that the CFT $(T^2)^n/\bbb{Z}_n$ is also a RCFT. Indeed, as we explain in detail in appendix \ref{kusuki}, we can show that the quantum dimension of the twist operator $\sigma_n$ in $(T^2)^n/\bbb{Z}_n$
takes the value
\be
\label{eq:QD3}
d_{\sigma_n}=\frac{S_{0\sigma_n}}{S_{00}}=\frac{1}{(s_{00})^{n-1}}=(2pq)^{n-1},
\ee
where $s_{00}=\frac{1}{2pq}$ is the vacuum S-matrix element in the $c=2$ CFT at the radius of $T^2$ equal $R=\sqrt{p/q}$. By using the formula  (\ref{qdim}), we can get the following expression of $\Delta S^{(2)}_A$:
\be
\label{eq:recover}
\Delta S^{(2)}_A=\log d_{\sigma_n}=(n-1)\cdot\log (2pq),
\ee
which perfectly matches (\ref{eq:Renyi}).
    \item For $n \to 1$, $\Delta S^{(2)}_A$ reduces to zero. This is consistent with the fact that the twist operators become the identity in the $n \to 1$ limit (no orbifold).
\end{itemize}

Note that, as shown in \cite{HNTW}, for RCFTs, the growth of $m$-th Renyi EE $\Delta S^{(m)}_A$ does not depend on $m$ and therefore we also expect the same for the rational $\eta$ in our setup
\be
\Delta S^{(m)}_A=(n-1)\cdot\log (2pq). \label{mRenyiRational}.
\ee
Moreover, we can extend the rational result into the case where we insert many twist operators.
As shown in \cite{Caputa2016yzn,Numasawa2016}, in RCFTs, entanglement is conserved
after the scattering between the local operators.
This means that the contribution from the local operators are summed up independently.
Therefore, when the excited state is created by insertion of $K$ twist operators,
we find the late time increase of the Renyi entropies
\be
\Delta S^{(m)}_A = K \cdot (n-1)\cdot\log (2pq).
\ee

\subsection{Evolution of Renyi Entanglement Entropy for Irrational $\eta$}
When $\eta$ is an irrational number, we simply find
\be
F_n(1,0)=0.  \label{wwe}
\ee
This means that the 2nd Renyi EE diverges with time. To see the exact time dependence we need to find a first order correction in the $\delta\to 0$ limit.\\
By using the result (\ref{eq:irrational}), we may naively estimate
\begin{equation}
\label{ntir}
F_n \PA{1-\frac{\epsilon^2}{4t^2},\frac{\epsilon^2}{4t^2}} \simeq\frac{2^{n-1}}{n} \cdot \delta^{n-1}=\frac{1}{n} \cdot \PA{\frac{2\pi}{\log (4t^2/\ep^2)}}^{n-1} .
\end{equation}
This leads to the following result of time evolution of $\Delta S^{(2)}_A$ when $\eta$ is irrational
\be
\Delta S^{(2)}_A\simeq (n-1)\cdot \log\left(\frac{\log (4t^2/\ep^2)}{2\pi}\right)+\log n \sim (n-1)\cdot \log\left(\log(t/\ep)\right).
\ee
This is proportional to the central charge $c=2n$ in the large $n$ limit and independent of $\eta$.

\begin{figure}[h!]
  \centering
  \includegraphics[width=7cm]{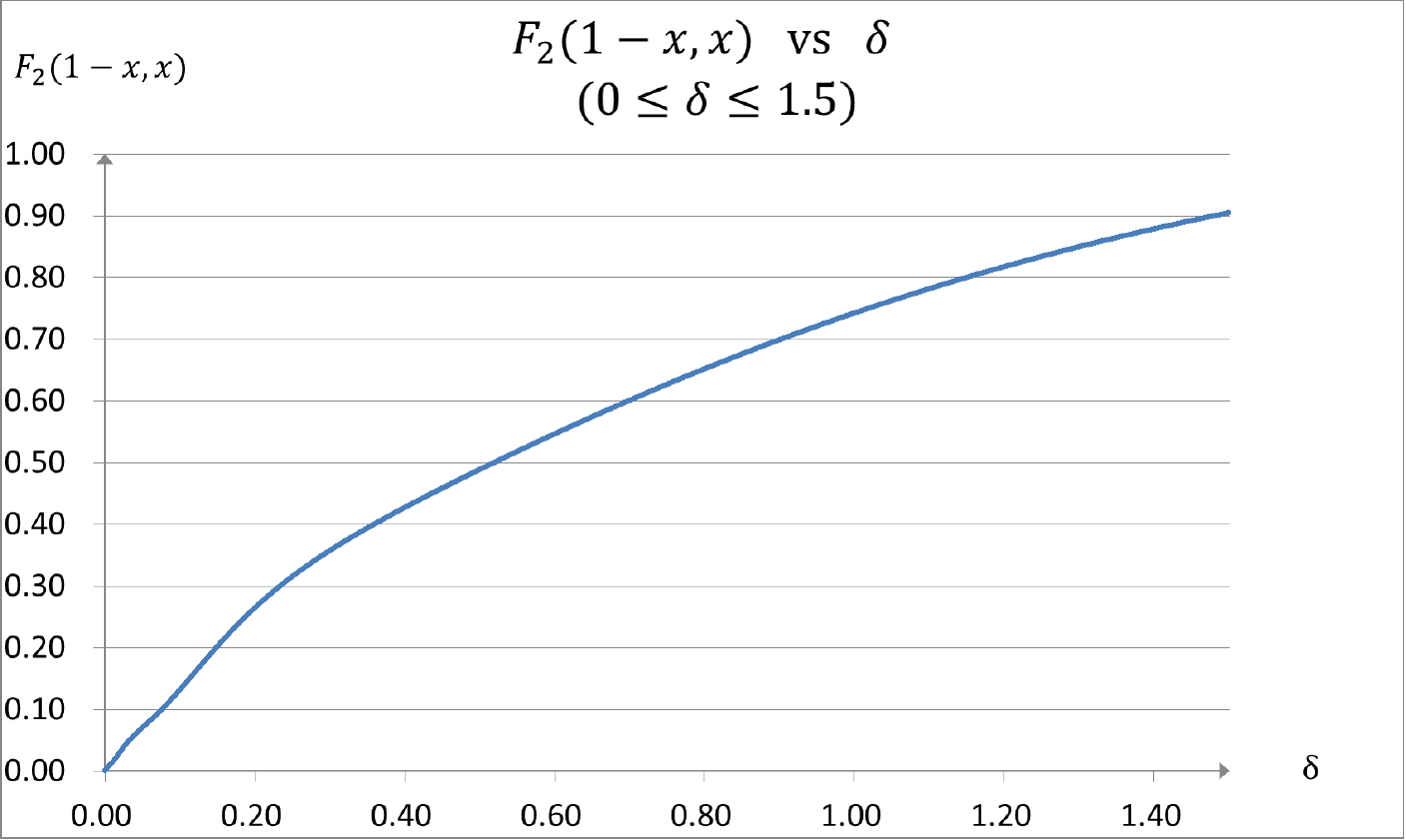}
  \includegraphics[width=7cm]{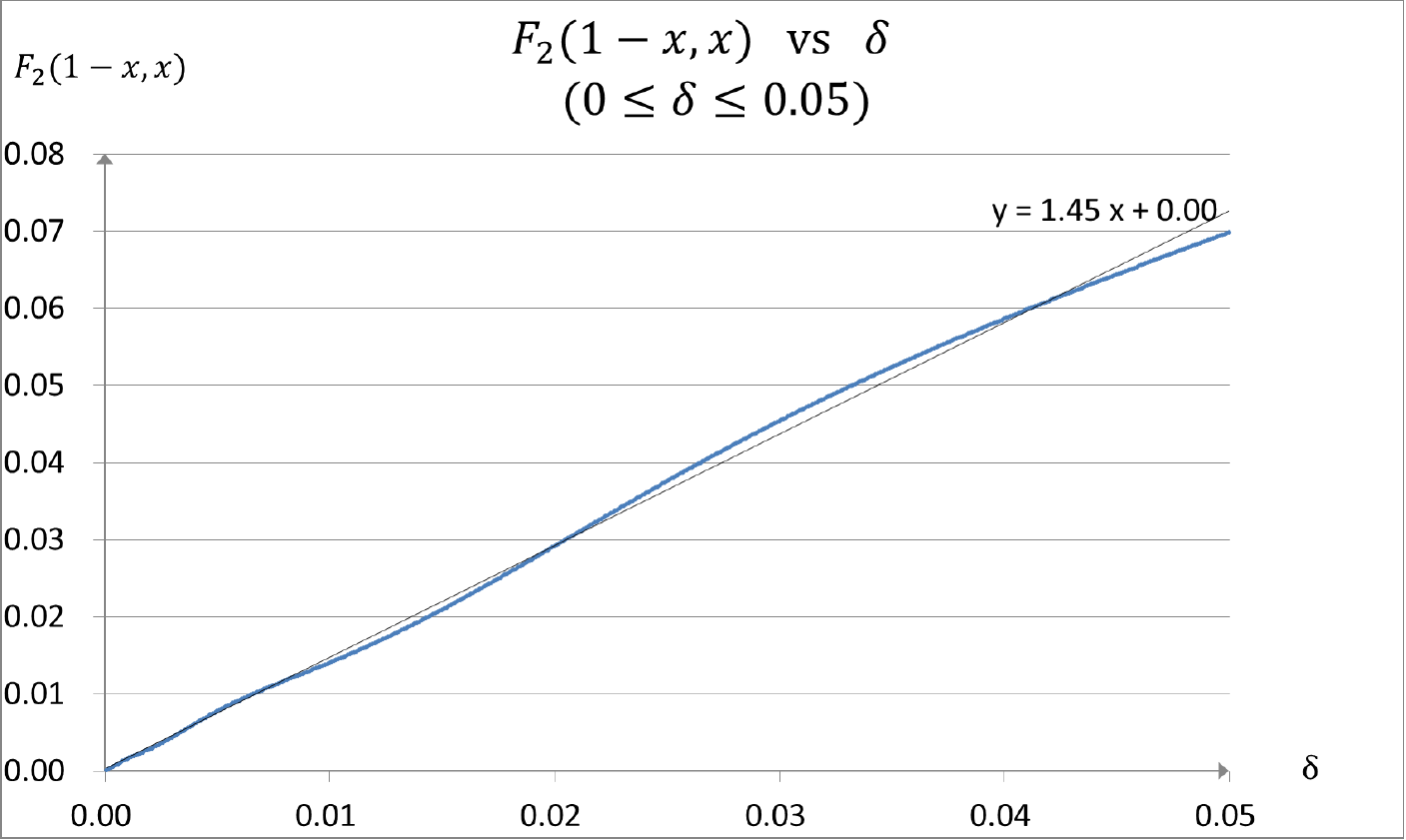}
  \caption{The plot of the function $F_2(x,1-x)$ for $\eta=\sqrt{2}$ as a function of $\delta=\frac{\pi}{-\log x}$. The left figure for $0\leq \delta\leq 1.6$. The right one for $0\leq \delta\leq 0.06$. The blue thick curves describe the plots. The black ones are the numerical fits. }\label{root2fig}
  \end{figure}

\begin{figure}[h!]
  \centering
  \includegraphics[width=7cm]{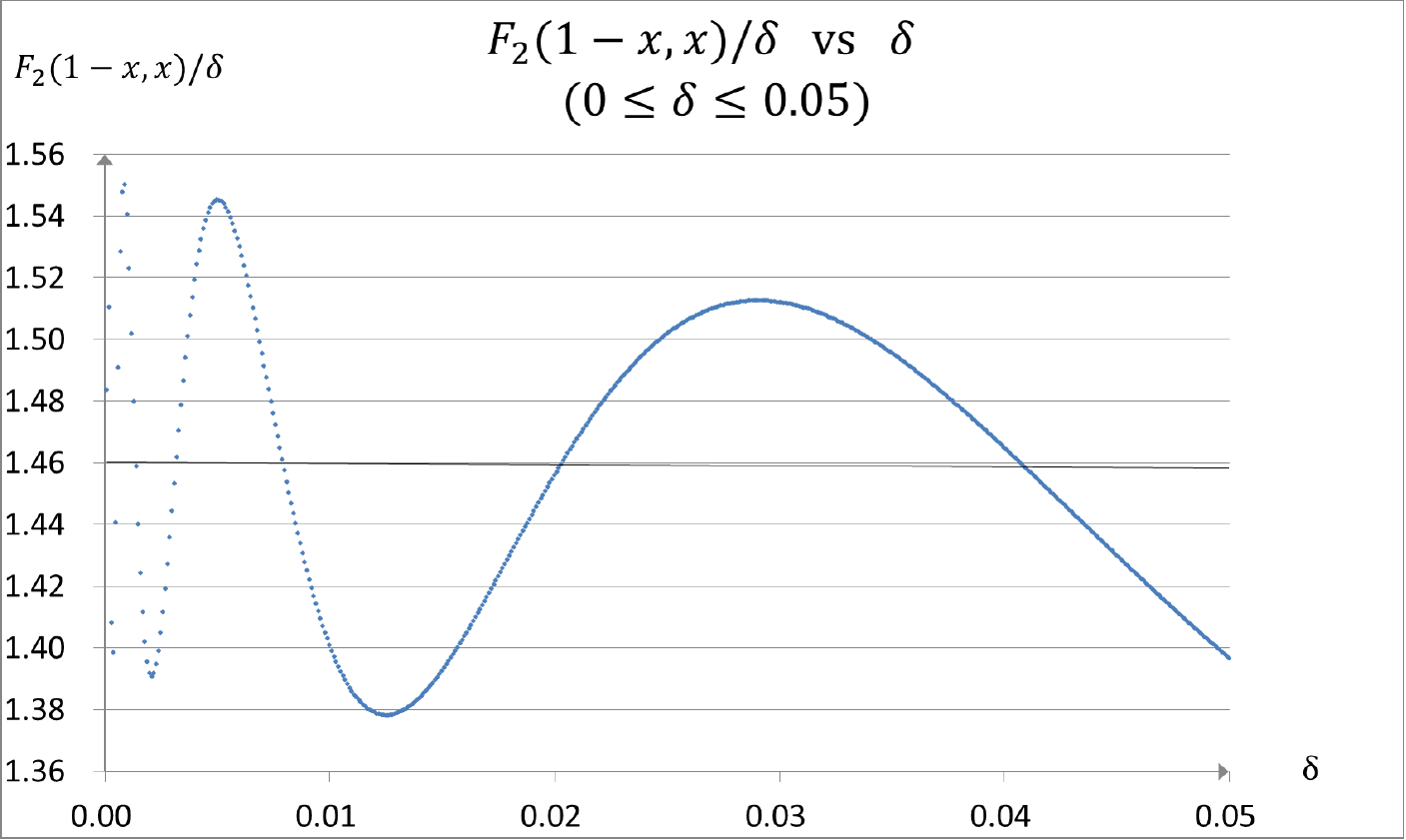}
  \caption{The plot of the ratio $F_2(x,1-x)/\delta$ for $\eta=\sqrt{2}$ as a function of $\delta=\frac{\pi}{-\log x}$. We took $0\leq \delta\leq 0.05$.}\label{root21fig}
  \end{figure}

However, strictly speaking, there is a subtle problem in this argument because we can make $|\ti{m}-\frac{\eta}{2} \Lambda_0 \cdot l|$ arbitrary small by taking $\ti{m}_r$ and $l_r$ large enough, even if $\eta$ is irrational. Since we do not have any analytical control on this problem, we performed numerical computations for various small values of $\delta$. The upshot is that the estimation (\ref{ntir})
is qualitatively correct. More explicitly, when $\delta$ is small, we find that the ratio
$F_n/\delta^{n-1}$ is bounded both from below and above
\be
\frac{2^{n-1}}{n}\leq \frac{F_n}{\delta^{n-1}} \leq A(\eta),
\ee
where $A(\eta)$ is a certain $O(1)$ constant which depends on $\eta$. Note that the lower bound is obvious because the following summation in (\ref{eq:gn})
\begin{equation}
\sum_{\bb{l},\bb{\ti{m}} \in {\bbb{Z}^{n-1}}}
e^{- \frac{\pi}{\eta \delta} (\bb{\ti{m}}^T - \frac{\eta}{2} \bb{l}^T \cdot\Lambda_0^T ) \cdot \Omega_0^{-1} \cdot (\bb{\ti{m}} -\frac{\eta}{2} \Lambda_0 \cdot \bb{l} ) -\pi \eta \delta \bb{l}^{T} \cdot \Omega_0 \cdot  \bb{l}}=1+\cdots
\end{equation}
 is larger than $1$. Indeed, in Fig.\ref{root2fig} and Fig.\ref{root21fig} we computed $F_2$ for $\eta=\s{2}$ numerically. It is clear that $F_2$  approaches to zero almost linearly like $F_2/\delta\sim 1.46$.\\
  
\begin{figure}
  \centering
  \includegraphics[width=7cm]{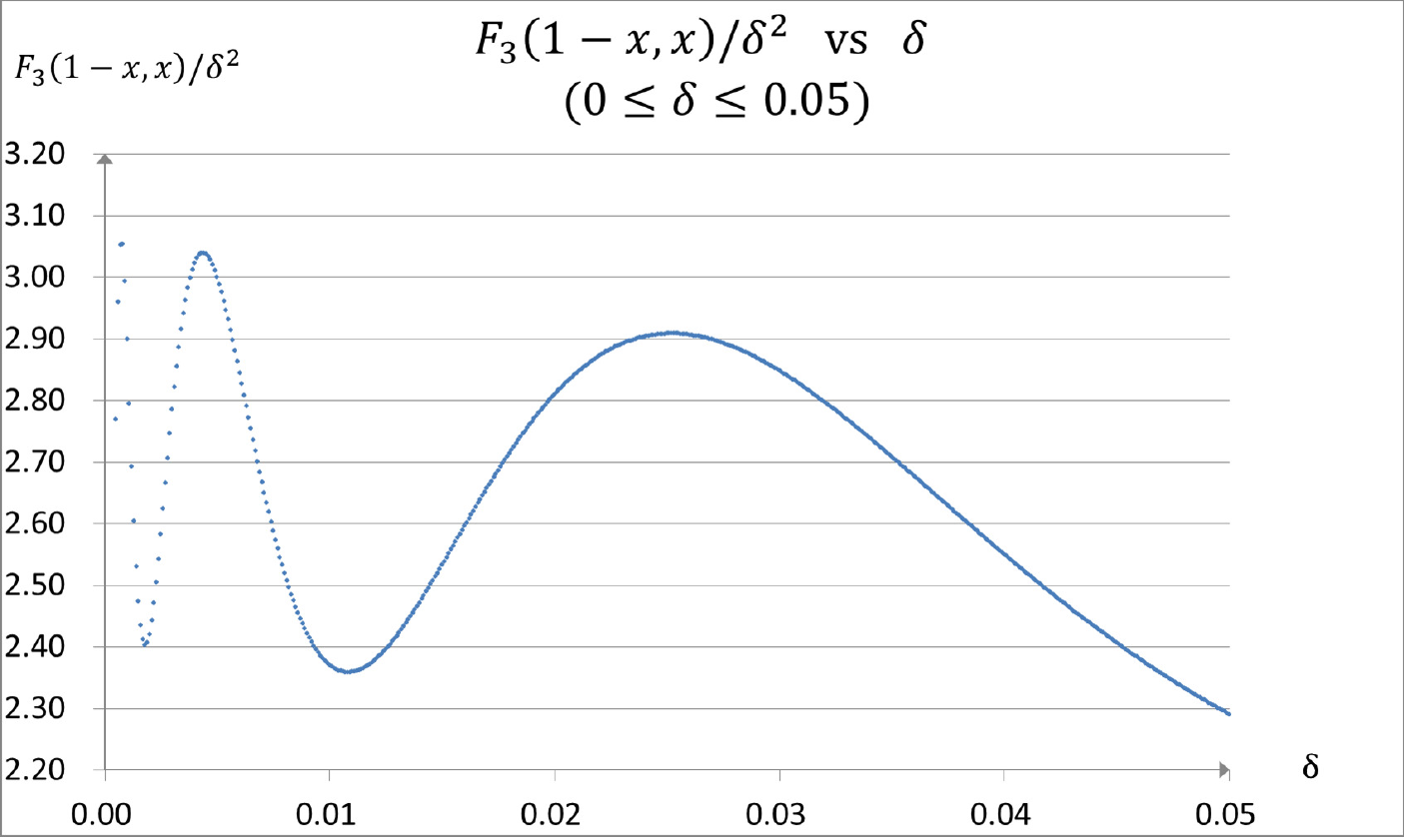}
  \caption{The plot of the ratio $F_3(x,1-x)/\delta^2$ for $\eta=\sqrt{2}$ as a function of $\delta=\frac{\pi}{-\log x}$. We took $0\leq \delta\leq 0.05$. }\label{root31fig}
  \end{figure}

Note also that Fig. \ref{root21fig} shows clear oscillations. This oscillation gets more frequent as we approache $\delta=0$. 
The reason why we have such oscillations is because we can approximate
any irrational number by infinitely many different rational numbers as in the continued fraction representation\footnote{For example, the continued fraction expansion of $\sqrt{2}\, ( = 1+\frac{1}{2+\frac{1}{2+ \cdots}} )$ generates
rational numbers $\{ \frac{p}{q} \} = 1, \frac{3}{2}, \frac{7}{5}, \frac{17}{12}, \frac{41}{29}, \cdots$ at each step.
This approximation causes the oscillation of $F_n$ locally minimized at $\delta \sim \frac{1}{2pq} = 0.5, 0.0833..., 0.0142..., 0.00245..., 0.000420..., \cdots$, as seen in Fig.\ref{root21fig} and Fig.\ref {root31fig}. However, for irrational numbers with complicated continued fraction expansion ($\pi, e, \dots$), it seems to be hard to see such structure.}, where $\delta$ measures the accuracy of the approximation.\\

\begin{figure}
  \centering
  \includegraphics[width=7cm]{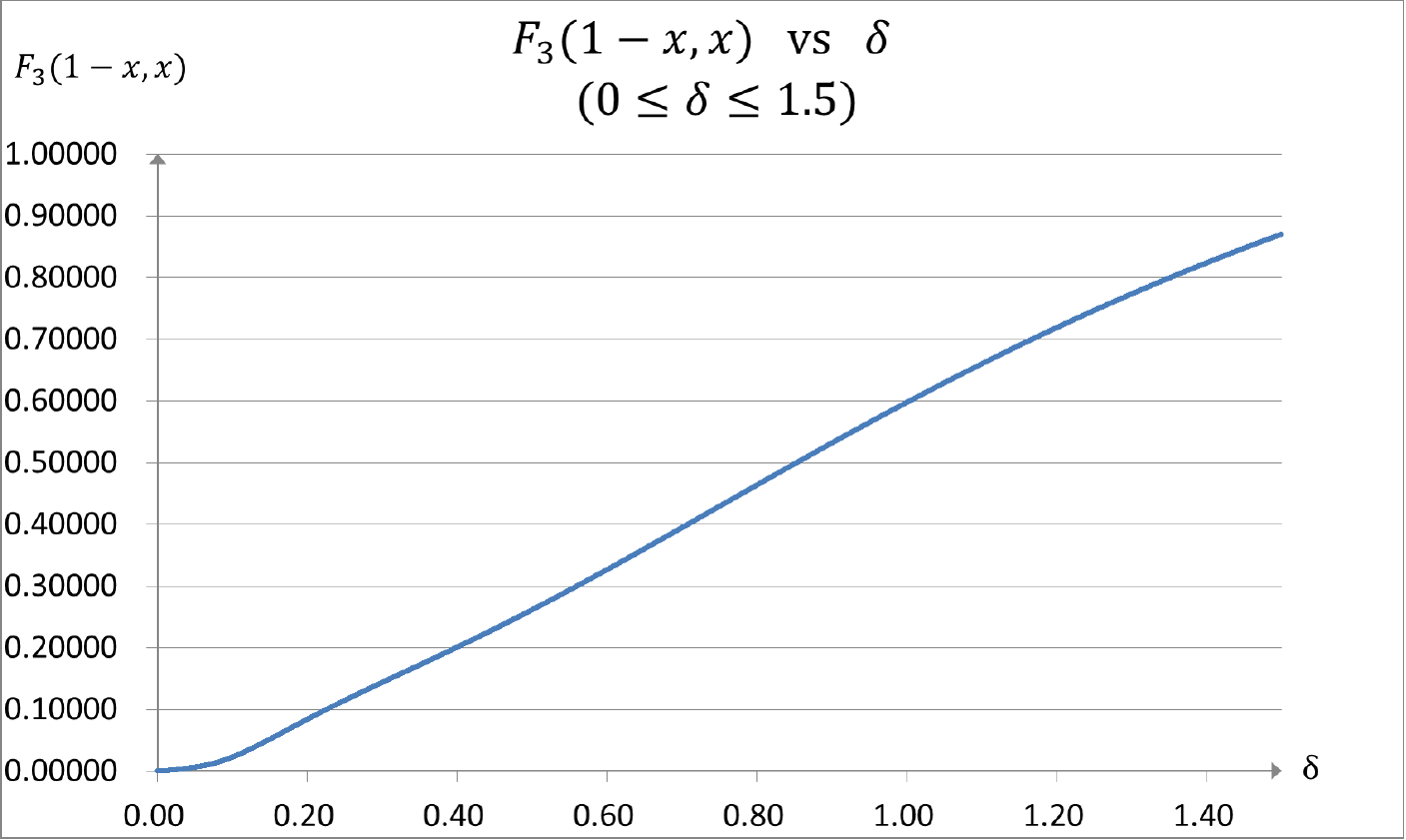}
  \includegraphics[width=7cm]{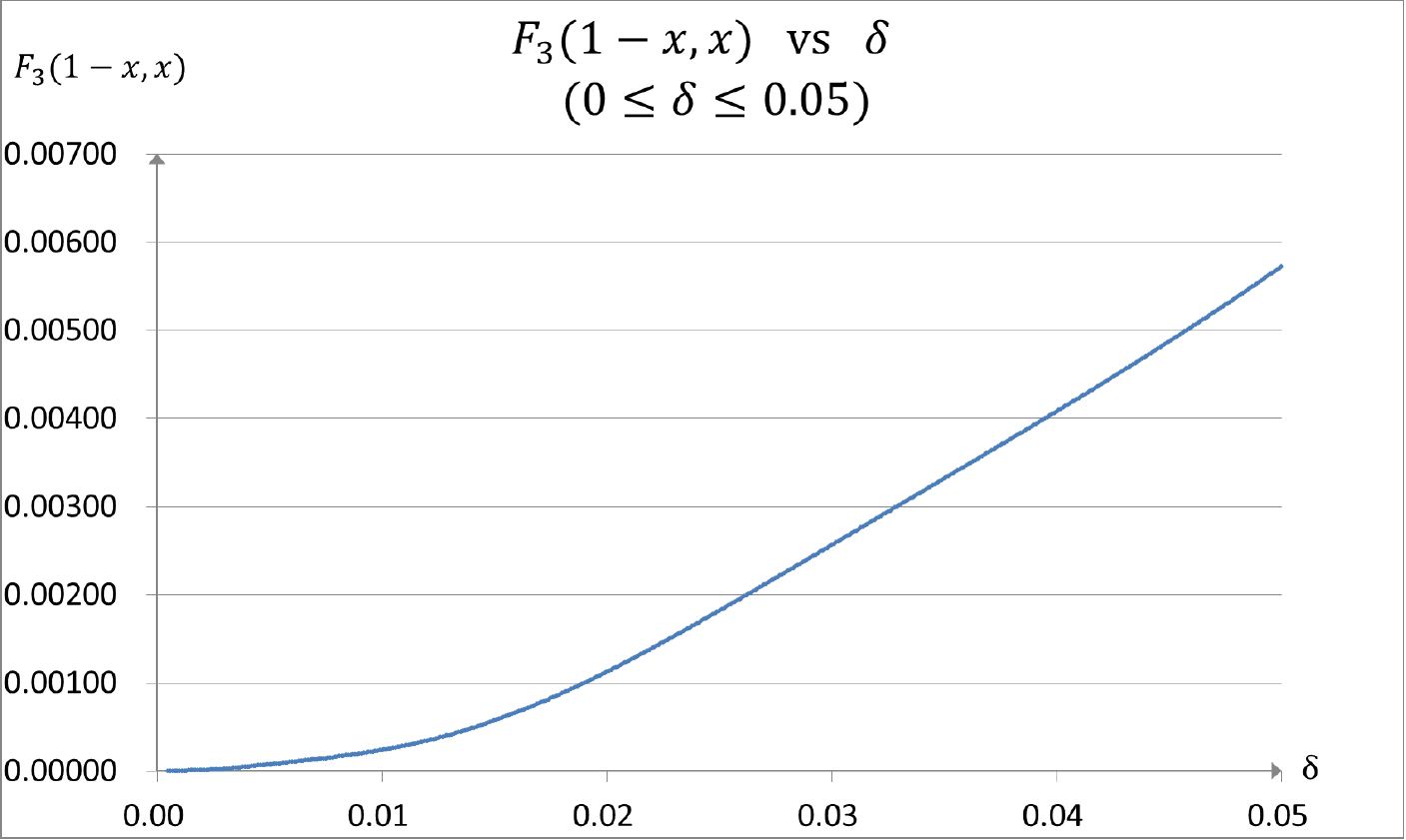}
  \caption{The plot of the function $F_3(x,1-x)$ for $\eta=\sqrt{2}$ as a function of $\delta=\frac{\pi}{-\log x}$. The left figure for $0\leq \delta\leq 1.6$. The right one for $0\leq \delta\leq 0.05$. The actual numerical plots are blue colored.}\label{root3fig}
  \end{figure}

In principle, we can extend our analysis to $n\geq 3$ in a straightforward way. However, our numerical analysis gets more involved as $n$ grows. We show our result for $n=3$ in Fig.\ref{root3fig} and Fig.\ref{root31fig} for $\eta=\s{2}$. The plots show that the ratio $F_3/\delta^2$ is bounded both from below and from above.

Summarizing, our thorough analysis shows that for irrational $\eta$, the growth of Renyi entanglement entropy has a form of the double logarithm 
\be
\Delta S^{(2)}_A\simeq (n-1)\cdot \log\left(\log(t/\ep)\right),
\ee
up to a constant term and ignoring the oscillating effect.

Note that, similarly to the rational case, it is natural to expected that $\Delta S^{(m)}_A$ also does not depend on $m$,
\be
\Delta S^{(m)}_A\simeq (n-1)\cdot \log\left(\log(t/\ep)\right), \label{mRenyiIrrational}
\ee
To show this intuitively, in the $\delta \to 0$ limit, we pinch off nontrivial cycles of the replica manifold as many as the genus $g = (m-1) (n-1)$. Then, $F_n$ obtains a divergent factor $\delta$ for each pinched cycle and the total divergent factor is $\delta^{g} = \delta^{(m-1)(n-1)}$.
This observation leads us to (\ref{mRenyiIrrational}). We can confirm it explicitly for the $m = 2$ case, as discussed in this section, however the cautious reader should bear in mind potential subtleties that could remain for irrational CFTs. It would be interesting to prove the independence on $m$ rigorously for irrational $\eta$ and we leave it as an open future problem.

\section{Growth of Renyi EE in Symmetric Orbifolds $(T^2)^n/S_n$}

So far we analyzed the Renyi EE for the cyclic orbifold $(T^2)^n/\bbb{Z}_n$. Now we would like to turn to the symmetric orbifold $(T^2)^n/S_n$ with the same radii $R$. Except $n=2$, they are different CFTs and any tractable formula of four point functions in the latter CFT is unfortunately not available at present. However, it is still correct that the symmetric orbifold CFT is rational if the value of $R^2$ is rational, which we set $R^2=p/q$ again. Therefore in this rational CFT case, we can apply (\ref{qqdim}) to calculate $\Delta S^{(m)}_A$. The detailed computation can be found in the appendix A. The upshot is that the quantum dimension takes the value
\be
d_{\sigma_n}=(n-1)!\cdot (2pq)^{n-1},
\ee
because we have an additional prefactor of $|C^A|=(n-1)!$ in (\ref{eq:QDn}) for the symmetric orbifold.
This leads to the growth of Renyi entropy
\be
\Delta S^{(m)}_A=(n-1)\log(2pq)+\log (n-1)!.  \label{dssym}
\ee

This result clearly shows the four point function of the twist operators in the cyclic orbifold CFT is different from that in the symmetric orbifold one. This difference is natural because the intermediate states in the computation of the four point function are projected by two different orbifold groups. The additional term in (\ref{dssym}) can be intuitively understood since the sizes of the two orbifold groups are different by the factor $|S_n|/|\bbb{Z}_n|=(n-1)!$. It will be an interesting future work to confirm our prediction (\ref{dssym}) by working out the four point function in $(T^2)^n/S_n$ explicitly and also compute the time evolution in the irrational case.

\section{Re-interpretation in terms of the Mutual Information}
\begin{figure}[h!]
  \centering
  \includegraphics[trim=100 100 100 100, width=5.5cm]{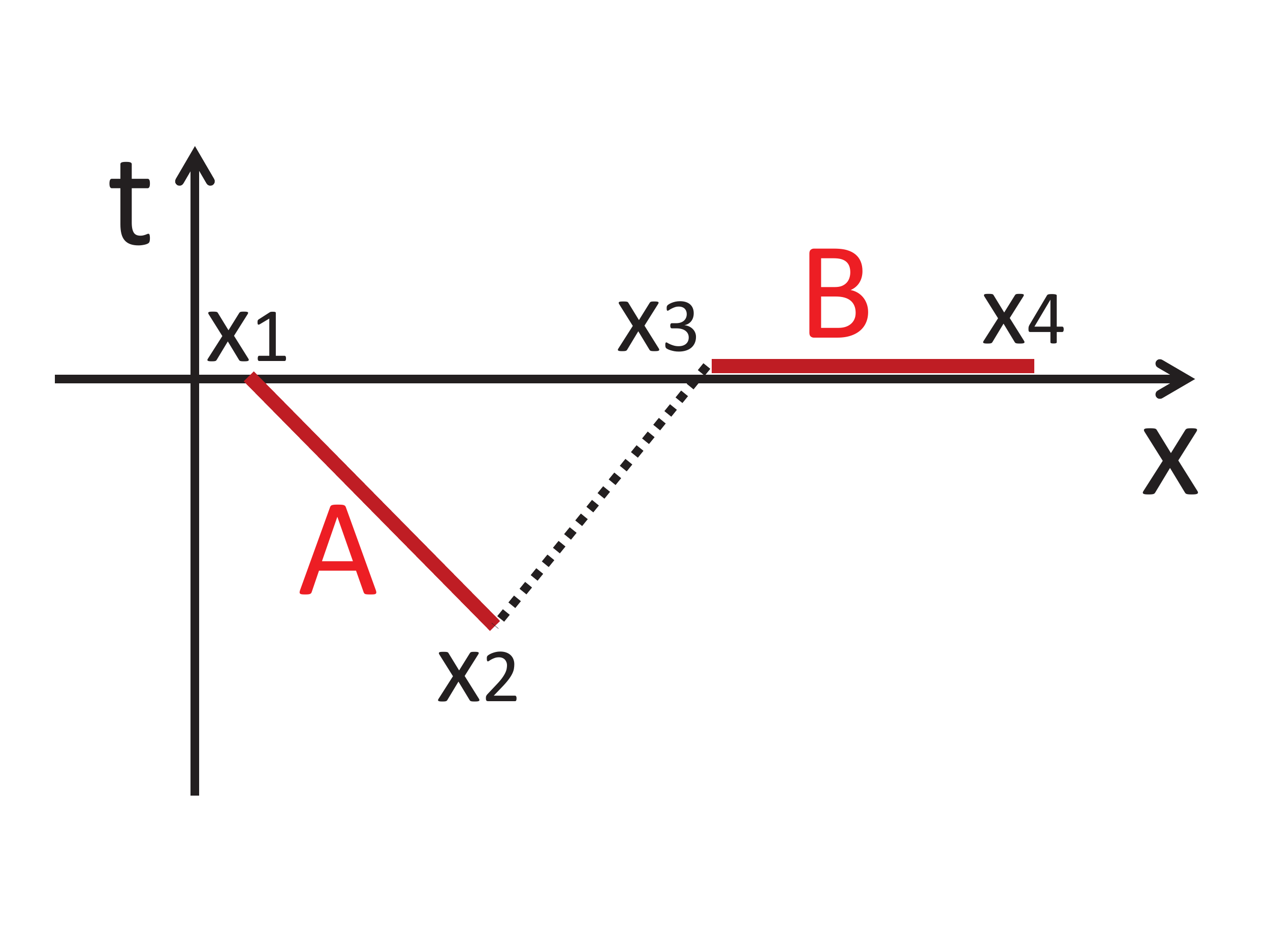}
  \caption{The setup of mutual information $I(A:B)$ between two intervals $A$ and $B$ when $A$ is infinitely boosted.}\label{merafig}
  \end{figure}

Since our analysis in the previous sections involves the four point function of twist operators, we can also interpret
the result in terms of the Renyi entanglement entropy $S^{(n)}_{A\cup B}$ for two intervals $A$ and $B$ as in \cite{Cardy,CCT,Headrick:2010zt,Hartman:2013mia},
or the Renyi mutual information
\be
 I^{(n)}(A : B)=S^{(n)}_{A}+S^{(n)}_{B}-S^{(n)}_{A\cup B} .
\ee
Let us choose $A$ and $B$ to be $[x_1,x_2]$ and $[x_3,x_4]$ in the two dimensional Lorentzian spacetime $\bbb{R}^{1,1}$.
Note that the Lorentzian time $t$ and space $x$ is related to the complex coordinate as $z=x+it_E=x-t$.
In terms of the complex coordinate, the intervals are specified by the twist operators at
\begin{align}
z_1 &= \bar{z}_1 =  x_1, \ \  z_3 = \bar{z}_3 = x_3,  \ \ z_4 = \bar{z}_4 = x_4, \notag \\
z_2 &= x_2 - (-t) , \ \ \bar{z}_2 = x_2 + (- t),
\end{align}
where $x_1 < x_2 < x_3 < x_4$ and $t > 0$.
Here we consider the simple case that only $x_2$ goes away from the $t = 0$ slice.
In this setup, the $(z,\bar{z})\to (1,0)$ limit corresponds to the "light-cone" limit.
In this limit, the interval $A$ is infinitely boosted and the Cauchy surface containing the intervals becomes singular (Fig.\ref{merafig}).\footnote{In \cite{Asplund2015a}, the light-cone limit of $S^{(n)}_{A \cup B}$ is discussed for globally excited states.
It detects the failure of the quasi-particle picture for propergation of entanglement.
Our setup is similar but different in that the excitation is (quasi-)locally caused by boosting the interval $A$. And, in \cite{Blanco:2011np}, other configurations of non-coplaner regions are discussed.}
To regularize this null limit, let us introduce regulators $\epsilon_{1,2}$ such that 
\be
x^{-}_{32}=-\ep_2(<0)\to 0,\ \ \ x^{+}_{21}=\ep_1(>0)\to 0.
\ee
where we defined the light-cone coordinate 
$x_j^\pm=t_j \pm x_j$ ($t_2=-t, t_1=t_3=t_4=0$).
The cross ratios are
\begin{align}
z
&= \frac{z_{12} z_{34}}{z_{13} z_{24}}
=\frac{x_{21}^{-} x_{43}^{-}}{x_{31}^{-} x_{42}^{-}}
= \frac{(x_{21} + t ) x_{43}}{ x_{31} (x_{42} - t)}
= 1 - \frac{ x_{32}^{-} x_{41}^{-}}{x_{31}^{-} (x_{43}^{-} + x_{32}^{-})}
= 1-\frac{(l+2t+\ep_1+\ep_2)\ep_2}{(2t+\ep_1+\ep_2)(l+\ep_2)}, \notag \\
\bar{z}
&= \frac{\bar{z}_{12} \bar{z}_{34}}{\bar{z}_{13} \bar{z}_{24}}
=\frac{x_{21}^{+} x_{43}^{+}}{x_{31}^{+} (x_{41}^{+} - x_{21}^{+} )}
= \frac{(x_{21} - t ) x_{43}}{ x_{31} (x_{42} + t )}
= \frac{l\ep_1}{(2t+\ep_1+\ep_2)(l+2t+\ep_2)}.
\end{align}
where $x_4-x_3=l$.
They are expanded in $\epsilon_{1,2}$ as follows
\begin{align}
z
\simeq  1-\left(\frac{l+2t}{2tl}\right)\cdot \ep_2 \cdots , \ \ \ \ \ \ 
\bar{z}
\simeq  \left(\frac{l}{2t(l+2t)}\right)\cdot \ep_1+ \cdots .
\end{align}
Especially, we focus on the expansions for $l \to \infty$ and $\ep_1 = \ep_2 = \ep$ 
which have quite simple forms 
as with (\ref{limep}).
\begin{align}
z
\simeq  1- \frac{\ep}{2t} \cdots , \ \ \ \ \ \ 
\bar{z}
\simeq \frac{\ep}{2t}+ \cdots .
\end{align}

Thus if we boost $A$ to almost null, then the $n$-th Renyi mutual information (\ref{eq:rational})
can be computed by performing the similar analysis for the $(z,\bar{z}) \to (1,0)$ limit of $F_n (z, \bar{z})$ in the previous sections.
 \be
 I^{(n)}(A:B) = \frac{1}{n-1} \log \left[ |1-z|^{-4 \Delta_{n}} \cdot F_n (z, \bar{z}) \right], \label{MI2.eq}
 \ee
where $\Delta_n =\frac{c}{24}(n-1/n)$.
For any $\eta$, the mutual information (\ref{MI2.eq}) has
the logarithmic divergent term coming from the first factor in the logarithm.
The $\eta$-dependence comes from $ F_n (z, \bar{z})$.
When $\eta=p/q$ is rational, applying (\ref{eq:rational}), we find the constant term.
\ba
I^{(n)}(A,B)=\frac{c}{12} \left(1+\frac{1}{n} \right) \log \left( \frac{y}{\epsilon_2} \right) -\log(2pq) ,
\ea
where $y=2t$.
We expect in any rational CFT this is generalized into the following form:
\ba
I^{(n)}(A,B)=\frac{c}{12} \left(1+\frac{1}{n} \right) \log \left( \frac{y}{\epsilon_2} \right) -\log d_{tot},
\ea
where $d_{tot}=1/s_{00}$ is the total quantum dimension of the (seed) CFT.
When $\eta$ is irrational,  applying (\ref{eq:irrational}),
we find the double logarithmic divergent term.
\ba
I^{(n)}(A:B)
= \frac{c}{12} \left(1+\frac{1}{n} \right) \log \left( \frac{y}{\epsilon_2} \right) - \log \left( \log \left( \frac{y}{\epsilon_2} \right) \right) - \frac{\log n}{n-1} + \log (2\pi).
\ea

\section{Conclusions}
In this work we studied the evolution of the Renyi entropy in cyclic and symmetric orbifold CFTs. As we showed, by considering twist operators as the primary excitations, we were able to extract various new and universal result from the four point correlators developed by Calabrese, Cardy, and Tonni \cite{CCT} for the cyclic orbifold CFT. Here we summarize our main findings and list some open problems.

Depending on the compactification radius, our setup can be divided into rational and irrational models. In the rational case the increase of the Renyi entropies approaches a universal constant at late times proportional to the logarithm of the total quantum dimension. This quantity is also defined as the inverse of the identity-identity component of the modular S-matrix of the seed theory. In our setup, for the rational radius $\eta=p/q$, we showed, both analytically and numerically, that for twist operators in the $n$-th cyclic orbifold CFT $(T^2)^n/\bbb{Z}_n$, the Renyi approaches the constant  $\Delta S^{(m)}=(n-1)\log (2pq)$. In the case of the symmetric orbifold CFT $(T^2)^n/S_n$, we get $\Delta S^{(m)}=(n-1)\log (2pq)+\log (n-1)!$. They do not agree with each other except $n=2$ because the intermediate states in the computation of four point functions are different due to different orbifold projections.

Moreover, we analyzed the time evolution of Renyi entropy in the cyclic orbifold CFT when the square of the radius is irrational. In this irrational case, we found a universal growth at late times given by the double logarithm of time. This is slower that the holographic one \cite{NNTH,CNT} but much faster and unconstrained that in RCFTs. This result strongly suggests that a breakdown of the quasi-particle picture, is strongly related to rationality of the underlying CFT. Motivated by our results we are tempting to conjecture that the
logarithmic growth found in the holographic CFTs is the fastest growth for the local excitations among all 2d CFTs. Such a systematical understanding of entanglement entropy growth
certainly deserves future studies.

Moreover, interestingly, our results for the second Renyi entropy can be interpreted in terms of the mutual information between light-like separated intervals. We derived a universal answer for the mutual information in this limit and it would be interesting to explore the physical meaning of this quantity in more details.

As another useful quantity which characterizes the evolutions of excitations is the out-of time order correlators (OTOCs), which provides a new tool to classify CFT from the perspective of the evolution of the quantum entanglement and information \cite{Roberts2015,Kitaev,Maldacena2016}.
 In our upcoming publication \cite{CKTW}, we will present the results for OTOCs for our symmetric orbifold CFTs and compare the results with those for RCFTs \cite{CNV,Gu2016} as well as chaotic CFTs .

\section*{ Acknowledgements}
We are grateful to Chris Herzog, Alvaro Veliz-Osorio, Tokiro Numasawa and Noburo Shiba for useful discussions and especially
to Erik Tonni for detailed explanations on relevant computations. PC is supported by the grant "Exact Results in Gauge and String Theories" from the Knut and Alice Wallenberg foundation. TT and PC are supported by the Simons Foundation through the ``It from Qubit'' collaboration. KW is supported by JSPS fellowshop. TT is supported by JSPS Grant-in-Aid for Scientific Research (A) No.16H02182 and World Premier International Research Center Initiative (WPI Initiative) from the Japan Ministry of Education, Culture, Sports, Science and Technology (MEXT). TT is also very grateful to the workshop ``Entanglement and Dynamical Systems'' held at Simons Center for Geometry and Physics, where this work was presented.

\appendix

\section{Quantum Dimension from S-matrices in Orbifold CFTs} \label{kusuki}
\subsection{Quantum dimension of twist operator}

If $\mathcal{C}$ denotes a RCFT, $\mathcal{C}^{\otimes n}$ denotes a $n$-th tensor product CFT of $\mathcal{C}$ and $\Omega$ denotes a permutation group in $S_n$,  the characters of the primary fields in $\mathcal{C}^{\otimes n}/\Omega$ and their modular property (i.e. S-matrix) are given in \cite{Bantay1998}. By using these expressions, we can get the quantum dimension of the twist operator in  $\mathcal{C}^{\otimes n}/\Omega$. To explain this, we use the following notation.

\hrulefill
\\
(Notation)\\
\\
$\mathcal{D}(\Omega)\cdots$The Drinfeld double of  the group $\Omega$, defined in \cite{Drinfeld1986}\\
\\
$p \cdots$ Some representative of an orbit of $\Omega$ acting on the $n$-tuples $\langle p_i,p_2,\ldots p_{n}\rangle$ of primaries $p_i$ of $\mathcal{C}$.\\
\\
$\phi \cdots$Irreducible character of the double $\mathcal{D}(\Omega_p)$ of the stabilizer $\Omega_p = \left\{x \in \Omega | xp = p \right\}$ of the $n$-tuple $p$\\
\\
$\langle p,\phi \rangle\cdots$The primary fields of $\mathcal{C}^{\otimes n}/\Omega$\\
\\
$\chi_p (\tau) \cdots$The genus one character of the primary field $p$ of $\mathcal{C}$.\\
\\
$\omega_p \cdots$The modular T-matrix of $\mathcal{C}$, in that $\omega_p =\mathrm{e}^{2 \pi i(\Delta_p-\frac{c}{24})}$\\
\\
$\mathcal{O}(x,y) \cdots$The set of orbits of the subgroup generated by $x$ and $y$, where a pair of $x,y \in \Omega$ is of commuting permutations\\
\\
To each ordered triple $\langle x, y, \xi \rangle$ with $\xi \in \mathcal{O}(x, y)$, we associate the following data:\\

1. $n_\xi$ ( resp. $n^*_\xi$ ) is the length of any $x$ orbit ( resp. $y$ orbit ) contained in $\xi$

2. $\mu_\xi$ ( resp. $\mu^*_\xi$) is the number of the $x$ orbits (resp. $y$ orbits )

3. $\kappa_\xi$ ( resp. $\kappa^*_\xi$) denotes the smallest non-negative integer for which $y^{\mu_\xi} =x^{\kappa_\xi}$( resp.  $y^{\mu^*_\xi} =x^{\kappa^*_\xi}$ ) holds on the points of $\xi$\\

\hrulefill

The characters of the primary fields $\langle p,\phi \rangle$ in $\mathcal{C}^{\otimes n}/\Omega$ are written as follows:
\begin{equation}
\label{eq:cha}
\chi_{\langle p,\phi \rangle}(\tau)=\frac{1}{|\Omega_p|}\sum_{x,y\in \Omega}\chi_p(x,y|\tau)\bar{\phi}(x,y),
\end{equation}
where

\begin{equation}
\chi_p(x,y|\tau)=\left\{
\begin{aligned}
\prod_{\xi \in \mathcal{O}(x,y)} &\omega_{p_{\xi}}^{-\frac{\kappa_{\xi}}{n_{\xi}}}\chi_{p_{\xi}}(\tau_{\xi})~~~&\text{if $x,y\in \Omega_p$ commute,}\\
&0 &\text{otherwise,}
\end{aligned}
\right.
\end{equation}
\begin{equation}
\tau_{\xi}=\frac{\mu_{\xi}\tau+\kappa_{\xi}}{n_{\xi}}
\end{equation}
and $p_{\xi}$ is the component of $p$ associated to the orbit $\xi$.
According to \cite{Bantay1998}, the conformal dimensions of the primary fields $\langle p,\phi \rangle$ can be obtained from the following relation:

\begin{equation}
\frac{1}{d_{\phi}} \sum_{x \in \Omega_p} \phi(x,x) \prod_{\xi \in \mathcal{O}(x,1)} \omega_{p_{\xi}}^{\frac{1}{|\xi|}}=\exp{\biggl(2 \pi i (\Delta_{\langle p,\phi \rangle}-\frac{n c}{24})\biggr)},
\end{equation}
where $\Delta_{\langle p,\phi \rangle}$ is the conformal dimension of the primary $\langle p,\phi \rangle$ of  $\mathcal{C}^{\otimes n}/\Omega$ and $d_{\phi}=\sum_{x \in \Omega_p} \phi(x,1)$. The irreducible character of the double $\phi$ can be labeled by a representative element $g^A$ of a conjugacy classes $C^A$ of $\Omega$ and an irreducible character $\alpha$ of the centralizer $Z_A$ of  $g^A$, so we denote by $\phi_{\alpha}^{g^A}$ the irreducible character of the quantum double.

Let's consider the special diagonal primary field $\langle p=(\Delta_p,\Delta_p,\cdots,\Delta_p),\phi_{\alpha}^{g^A}\rangle$ where $g^A$ is the cyclic permutation $(1,2,\cdots,n)$ and $\alpha$ is the trivial irreducible representation of $Z_A$. In this case,

\begin{equation}
\label{eq:double}
\phi_{\alpha}^{g^A}(x,y)=\left\{
\begin{aligned}
&0,~~~\text{if } x \notin C^A ~\text{or } xy \neq yx\\
&1,~~~\text{othewise}
\end{aligned}
\right.
\end{equation}
and

\begin{equation}
\begin{aligned}
\frac{1}{d_{\phi}} \sum_{x \in \Omega_p} \phi(x,x) \prod_{\xi \in \mathcal{O}(x,1)} \omega_{p_{\xi}}^{\frac{1}{|\xi|}} &=\prod_{\xi \in \mathcal{O}(g^A,1)} \omega_{p_{\xi}}^{\frac{1}{|\xi|}}\\
&=\omega_{p_{\xi}}^{\frac{1}{n}}\\
&=\Exp{2 \pi i(\frac{\Delta_p}{n}-\frac{c}{24 n})}\\
&=\exp{\biggl(2 \pi i (\Delta_{\langle p,\phi \rangle}-\frac{n c}{24})\biggr)}.
\end{aligned}
\end{equation}
Therefore the conformal dimension of $\langle p=(\Delta_p,\Delta_p,\cdots,\Delta_p),\phi_{\alpha}^{g^A}\rangle$ is

\begin{equation}
\Delta_{\langle p,\phi \rangle}=\frac{\Delta_p}{n}+\frac{c}{24}(n-\frac{1}{n}).
\end{equation}
In particular,
\begin{equation}
\Delta_{\langle 0,\phi \rangle}=\frac{c}{24}(n-\frac{1}{n}),
\end{equation}
where $0$ is the set of vacuum, $(0,0,\cdots,0)$. This is exactly the same as the weight of the twist operator.

Next, let's calculate the special element of S-matrix, $S_{0,\langle 0,\phi_{\alpha}^{g^A} \rangle}$. In general, the elements of the S-matrix of $\mathcal{C}^{\otimes n}/\Omega$ is very complicated, but the special elements, $S_{0,\langle p, \phi \rangle}$, have the following simple form:

\begin{equation}
\label{eq:S}
S_{0,\langle p, \phi \rangle}=\frac{1}{|\Omega_p|} \sum_{x \in \Omega_p} \phi(x,1) \prod_{\xi \in \mathcal{O}(x,1)} S^{\text{seed}}_{0 p_\xi},
\end{equation}
where $S^{\text{seed}}$ is the S-matrix of  $\mathcal{C}$. Inserting (\ref{eq:double}) into (\ref{eq:S}), we can get this element, in that,

\begin{equation}
\begin{aligned}
S_{0,\langle 0,\phi_{\alpha}^{g^A} \rangle}
&=\frac{|C^A|}{|\Omega|}\prod_{\xi \in \mathcal{O}(g^A,1)} S^{\text{seed}}_{0 0_\xi}\\
&=\frac{|C^A|}{|\Omega|}(S^{\text{seed}}_{00}).
\end{aligned}
\end{equation}
On the other hand, the trivial character of the quantum double is
\begin{equation}
\phi_{\alpha}^{e}(x,y)=\delta_{e,x}.
\end{equation}
By the same calculation as the above, we can get
\begin{equation}
S_{0,0}=\frac{1}{|\Omega|}(S^{\text{seed}}_{00})^{n},
\end{equation}
hence the quantum dimension $d_{\sigma_n}$ of the twist operator is
\begin{equation}
\label{eq:QDn}
d_{\sigma_n}=\frac{S_{0,\langle 0,\phi_{\alpha}^{g^A} \rangle}}{S_{0,0}}=\frac{|C^A|}{(S^{\text{seed}}_{00})^{n-1}}.
\end{equation}

This result holds in general RCFTs, in that the quantum dimension of the twist operator in $(\text{general RCFT})^{\otimes n}/\Omega$ is related to the ($n-1$)-th power of the total dimension of its seed RCFT. But as it is obvious from the above derivation, $\Omega$ have to contain the group element $(1,2,\cdots,n)$. $\bbb{Z}_{n}$ has the cyclic permutation $(1,2,\cdots,n)$, so we can use the above formula (in this case, $|C^A|=1$).

Let's focus on our case where $\mathcal{C}$ is $T^2$ with $\eta=p/q$. As is well known, the characters of the primary fields in the free compactified boson theory with  $\eta=p/q$  are written as follows:
\begin{equation}
\begin{aligned}
\chi^{(pq)}_l (q)&=\frac{1}{\eta (q)}\sum_{m \in \bbb{Z}} q^{\frac{1}{4pq}(l+2 m p q)^2}\\
&=\frac{\Theta_{l,pq}(\tau,0)}{\eta (\tau)},~~~-pq+1\leq l \leq pq,
\end{aligned}
\end{equation}
where $q=\mathrm{e}^{2 \pi i \tau}$, $\Theta_{l,k}(\tau,z)$ are the theta functions and $\eta(\tau)$ is the eta function. Using the modular transformation law of the theta functions and the eta function, we can get the following modular properties:
\begin{equation}
\chi^{(pq)}_l \biggl(-\frac{1}{\tau}\biggr)=\sum_{m=-pq+1}^{pq} \frac{\mathrm{e}^{\frac{-\pi i l m}{pq}}}{\sqrt{2pq}} \chi^{(pq)}_m (\tau).
\end{equation}
Therefore the modular S-matrix of this theory is
\begin{equation}
\label{eq:Su}
S^{T^1}_{lm}=\frac{\mathrm{e}^{\frac{-\pi i l m}{pq}}}{\sqrt{2pq}}.
\end{equation}
We can get the modular matrix element $S^{T^2}_{00}$ of $T^2$ just by squaring $S^{T^1}_{00}$, so we get
\begin{equation}
S^{\text{seed}}_{00}=S^{T^2}_{00}=\frac{1}{2pq}.
\end{equation}
Using (\ref{eq:QDn}), the quantum dimension of the twist operator in $(T^2)^n/\bbb{Z}_n$ is
\begin{equation}
\label{eq:d}
d_{\sigma_n}=(2pq)^{n-1},
\end{equation}
which is exactly (\ref{eq:QD3}) we wanted to prove.
Note that, from the above derivation, we can understand that $2pq$ reflects the number of irreducible characters.

\subsection{Application and Another calculation for $S_n$}
We showed the quantum dimension for  $(T^2)^n/\bbb{Z}_n$ by (\ref{eq:d}) in order to recover (\ref{eq:recover}), but it's also interesting to consider $S_n$ case as we said in the introduction. From (\ref{eq:QDn}), we can directly obtain the quantum dimension of twist operator for $S_n$:
\begin{equation}
d_{\sigma_n}=\frac{(n-1)!}{(S^{\text{seed}}_{00})^{n-1}},
\end{equation}
where we use the fact that $|C_A|=(n-1)!$ for $g^A=(1,2,3,\cdots,n)$ in $S_n$.

In fact, we can check our result for $S_n$ case from the another viewpoint. If we have the explicit expressions for the characters of the twist and vacuum operator, we can evaluate the quantum dimension of the twist field in the following way:

\begin{equation}
\label{eq:limitQD}
\begin{aligned}
\lim_{\tau \to i\infty}\frac{\chi_i\PA{-\frac{1}{\tau}}}{\chi_0 \PA{-\frac{1}{\tau}}}&=\lim_{\tau \to i\infty}\frac{\sum_j S_{ij}\chi_j(\tau)}{\sum_j S_{0j}\chi_j(\tau)}\\
&=\frac{S_{i0}}{S_{00}}\\
&=d_i.
\end{aligned}
\end{equation}
The characters of $S_3$ orbifolds has been presented  in \cite{JevickiYoon2016} and in particular, the characters of the vacuum and twist operator are as follows:

\begin{table}[!h]
\-\hspace{-2em}
{\renewcommand{\arraystretch}{2}
\begin{tabular}{|>{\centering} m{2.6cm}|>{\centering} m{9.3cm}|>{\centering} m{2.3cm} |}
\hline
Meaning\\ if $h_p=0$ & $S_3$ orbifold character & Dimension   \tabularnewline \hline
vacuum & $\frac{1}{6}\left[\left(\chi\left(\tau\right)\right)^3+3\chi\left(\tau\right)\chi\left(2\tau\right)+2\chi\left(3\tau\right)\right]$ & $3h_p$\tabularnewline \hline
twist op. & $\frac{1}{3}\left[\chi\left(\frac{\tau}{3}\right)+\chi\left(\frac{\tau+1}{3}\right)+\chi\left(\frac{\tau+2}{3}\right)\right]$ & $\frac{1}{3}h_p+\frac{1}{9}c$    \tabularnewline\hline
\end{tabular}
}
\caption{Characters for primaries $\langle p,p,p\rangle$ of $S_3 $ orbifold.}
\end{table}

If the character of the orbifold denotes $\mathcal{X}_i$, the characters corresponding to the vacuum ($0$) and twist ($\sigma$) operator are respectively

\begin{equation}
\mathcal{X}_0(\tau)=\frac{1}{6}\left[\left(\chi_0\left(\tau\right)\right)^3+3\chi_0\left(\tau\right)\chi_0\left(2\tau\right)+2\chi_0\left(3\tau\right)\right]
\end{equation}
and
\begin{equation}
\mathcal{X}_{\sigma}(\tau)=\frac{1}{3}\left[\chi_0\left(\frac{\tau}{3}\right)+\chi_0\left(\frac{\tau+1}{3}\right)+\chi_0\left(\frac{\tau+2}{3}\right)\right].
\end{equation}

To apply these to (\ref{eq:limitQD}), we need to evaluate the limit of the following quantities.

\begin{equation}
\begin{aligned}
\chi_i\PA{-\frac{p}{\tau}}&=\sum_j S_{ij} \chi_j \PA{\frac{\tau}{p}}\\
&\xrightarrow{\tau\to i\infty}S_{i0}\PA{q^{-\frac{c}{24}}}^\frac{1}{p},
\end{aligned}
\end{equation}
\begin{equation}
\begin{aligned}
\chi_i\PA{\frac{-\frac{1}{\tau}+1}{3}}&=\sum_j S_{ij} \chi_j \PA{\frac{3\tau}{1-\tau}}\\
&=\sum_j S_{ij} \chi_j \PA{-3+\frac{3}{1-\tau}}\\
&=\sum_j (ST^{-3})_{ij} \chi_j \PA{\frac{3}{1-\tau}}\\
&=\sum_j (ST^{-3}S)_{ij} \chi_j \PA{\frac{\tau-1}{3}}\\
&\xrightarrow{\tau\to i\infty}(ST^{-3}S)_{i0}\PA{q^{-\frac{c}{24}}}^\frac{1}{3}\times (\mathrm{e}^{-\frac{2 \pi i}{3}})^{-\frac{c}{24}}
\end{aligned}
\end{equation}
and
\begin{equation}
\begin{aligned}
\chi_i\PA{\frac{-\frac{1}{\tau}+2}{3}}&=\sum_j T_{ij}\chi_j\PA{\frac{-\frac{1}{\tau}-1}{3}}\\
&=\sum_j (TST^{3}S)_{ij} \chi_j \PA{\frac{\tau+1}{3}}\\
&\xrightarrow{\tau\to i\infty}(TST^{3}S)_{i0}\PA{q^{-\frac{c}{24}}}^\frac{1}{3}\times (\mathrm{e}^{\frac{2 \pi i}{3}})^{-\frac{c}{24}}.
\end{aligned}
\end{equation}

By using these, the leading term of $\mathcal{X}_{0}(-\frac{1}{\tau})$ ($\tau \to i\infty$) is
\begin{equation}
\mathcal{X}_0\PA{-\frac{1}{\tau}}\xrightarrow{\tau\to i\infty}\frac{1}{6}S_{00}^3\PA{q^{-\frac{c}{24}}}^3
\end{equation}
and that of $\mathcal{X}_{\sigma}(-\frac{1}{\tau})$ is
\begin{equation}
\mathcal{X}_{\sigma}\PA{-\frac{1}{\tau}}\xrightarrow{\tau\to i\infty}\frac{1}{3}S_{00}\PA{q^{-\frac{c}{24}}}^3.
\end{equation}
Inserting these into (\ref{eq:limitQD}),

\begin{equation}
d_{\sigma}=2\frac{1}{S_{00}^2}.
\end{equation}

In the same way, we can also calculate the quantum dimension for $S_4$ case.

\begin{table}[h]
\-\hspace{-2em}
{\renewcommand{\arraystretch}{2}
\begin{tabular}{|>{\centering} m{2.6cm}|>{\centering} m{9.3cm}|>{\centering} m{2.3cm} |}
\hline
Meaning\\ if $h_p=0$ & $S_4$ orbifold character & Dimension   \tabularnewline \hline
vacuum & $\frac{1}{4!}\left[\left(\chi(\tau)\right)^4+6\chi(2\tau)\left(\chi(\tau)\right)^2\right.$ $\left.+8\chi(3\tau)\chi(\tau)+3\left(\chi(2\tau)\right)^2+6\chi(4\tau)\right]$ & $4h_p$\tabularnewline \hline
twist op. & $\frac{1}{4}\left[\chi\left(\frac{\tau}{4}\right)+\chi\left(\frac{\tau+1}{4}\right)+\chi\left(\frac{\tau+2}{4}\right)+\chi\left(\frac{\tau+3}{4}\right)\right]$ & $\frac{1}{4}h_p+\frac{5}{32}c$    \tabularnewline\hline
\end{tabular}
}
\caption{Characters for primaries $\langle p,p,p,p\rangle$ of $S_4 $ orbifold.}
\end{table}

According to the above table, the leading term of $\mathcal{X}_{0}(-\frac{1}{\tau})$ ($\tau \to i\infty$) is

\begin{equation}
\mathcal{X}_0\PA{-\frac{1}{\tau}}\xrightarrow{\tau\to i\infty}\frac{1}{4!}S_{00}^4\PA{q^{-\frac{c}{24}}}^4
\end{equation}
and that of $\mathcal{X}_{\sigma}(-\frac{1}{\tau})$ is
\begin{equation}
\mathcal{X}_{\sigma}\PA{-\frac{1}{\tau}}\xrightarrow{\tau\to i\infty}\frac{1}{4}S_{00}\PA{q^{-\frac{c}{24}}}^4.
\end{equation}
Inserting these into (\ref{eq:limitQD}),

\begin{equation}
d_{\sigma}=3!\frac{1}{S_{00}^3}.
\end{equation}

The analysis of the higher $n$ proceeds along the same lines and the expression for general $n$ is supposed as follows: 
\begin{equation}
d_{\sigma_n}=\frac{(n-1)!}{(S^{\text{seed}}_{00})^{n-1}}.
\end{equation}
This result is consistent with our general expression.

\section{The determinant of $\Omega_0$} \label{kusuki2}
The characteristic polynomial of $\Omega_0$ is
\begin{equation}
|\Omega_0-\lambda|=(\frac{4}{n}-\lambda)(4-\lambda)^{n-2}=0,
\end{equation}
where we insert (\ref{eq:NMat}) into the following formula for $(n-1) \times (n-1)$ matrices:

\begin{equation}
  \left|
    \begin{array}{ccccc}
      a & b & b& \cdots &  b \\
      b & a & b & \cdots & b \\
      b& b & a &\cdots &  b \\
      \vdots &\vdots & \vdots & \ddots& \vdots \\
      b&  b & b &\cdots &  a
    \end{array}
  \right|=(a+(n-2)b)(a-b)^{n-2}.
\end{equation}
Therefore the eigenvalues of $\Omega_0^{-1}$ are $\{ \frac{n}{4},\frac{1}{4},\frac{1}{4},\ldots\}$ and these are clearly positive. And we can also evaluate the determinant of $\Omega_0$ as follows:
\begin{equation}
\det \Omega_0=\frac{4^{n-1}}{n},
\end{equation}
which is exactly (\ref{eq:Omega}) we wanted to prove.

\bibliographystyle{JHEP-2}
\bibliography{SP}

\providecommand{\href}[2]{#2}\begingroup\raggedright\begin{thebibliography}{10}

\bibitem{Cardy:1986ie}
J.~L. Cardy, {\it {Operator Content of Two-Dimensional Conformally Invariant
  Theories}},  {\em Nucl. Phys.} {\bf B270} (1986) 186--204.

\bibitem{HLW}
C.~Holzhey, F.~Larsen and F.~Wilczek, {\it {Geometric and renormalized entropy
  in conformal field theory}},  {\em Nucl. Phys.} {\bf B424} (1994) 443--467
  [\href{http://arXiv.org/abs/hep-th/9403108}{{\tt hep-th/9403108}}].

\bibitem{Cardy}
P.~Calabrese and J.~L. Cardy, {\it {Entanglement entropy and quantum field
  theory}},  {\em J. Stat. Mech.} {\bf 0406} (2004) P06002
  [\href{http://arXiv.org/abs/hep-th/0405152}{{\tt hep-th/0405152}}].

\bibitem{GQ}
P.~Calabrese and J.~L. Cardy, {\it {Evolution of entanglement entropy in
  one-dimensional systems}},  {\em J. Stat. Mech.} {\bf 0504} (2005) P04010
  [\href{http://arXiv.org/abs/cond-mat/0503393}{{\tt cond-mat/0503393}}].

\bibitem{CCL}
P.~Calabrese and J.~L. Cardy, {\it {Time-dependence of correlation functions
  following a quantum quench}},  {\em Phys. Rev. Lett.} {\bf 96} (2006) 136801
  [\href{http://arXiv.org/abs/cond-mat/0601225}{{\tt cond-mat/0601225}}].

\bibitem{CCR}
P.~Calabrese and J.~Cardy, {\it {Entanglement entropy and conformal field
  theory}},  {\em J. Phys.} {\bf A42} (2009) 504005
  [\href{http://arXiv.org/abs/0905.4013}{{\tt 0905.4013}}].

\bibitem{CCR2}
P.~Calabrese and J.~Cardy, {\it {Quantum quenches in 1+1 dimensional conformal
  field theories}},  {\em J. Stat. Mech.} {\bf 1606} (2016), no.~6 064003
  [\href{http://arXiv.org/abs/1603.02889}{{\tt 1603.02889}}].

\bibitem{RT1}
S.~Ryu and T.~Takayanagi, {\it {Holographic derivation of entanglement entropy
  from AdS/CFT}},  {\em Phys. Rev. Lett.} {\bf 96} (2006) 181602
  [\href{http://arXiv.org/abs/hep-th/0603001}{{\tt hep-th/0603001}}].

\bibitem{RT2}
S.~Ryu and T.~Takayanagi, {\it {Aspects of Holographic Entanglement Entropy}},
  {\em JHEP} {\bf 08} (2006) 045
  [\href{http://arXiv.org/abs/hep-th/0605073}{{\tt hep-th/0605073}}].

\bibitem{RT3}
V.~E. Hubeny, M.~Rangamani and T.~Takayanagi, {\it {A Covariant holographic
  entanglement entropy proposal}},  {\em JHEP} {\bf 07} (2007) 062
  [\href{http://arXiv.org/abs/0705.0016}{{\tt 0705.0016}}].

\bibitem{RTR1}
T.~Nishioka, S.~Ryu and T.~Takayanagi, {\it {Holographic Entanglement Entropy:
  An Overview}},  {\em J. Phys.} {\bf A42} (2009) 504008
  [\href{http://arXiv.org/abs/0905.0932}{{\tt 0905.0932}}].

\bibitem{RTR2}
T.~Takayanagi, {\it {Entanglement Entropy from a Holographic Viewpoint}},  {\em
  Class. Quant. Grav.} {\bf 29} (2012) 153001
  [\href{http://arXiv.org/abs/1204.2450}{{\tt 1204.2450}}].

\bibitem{RTR3}
M.~Rangamani and T.~Takayanagi, {\it {Holographic Entanglement Entropy}},
  \href{http://arXiv.org/abs/1609.01287}{{\tt 1609.01287}}.

\bibitem{AAL}
J.~Abajo-Arrastia, J.~Aparicio and E.~Lopez, {\it {Holographic Evolution of
  Entanglement Entropy}},  {\em JHEP} {\bf 11} (2010) 149
  [\href{http://arXiv.org/abs/1006.4090}{{\tt 1006.4090}}].

\bibitem{Ba}
V.~Balasubramanian, A.~Bernamonti, J.~de~Boer, N.~Copland, B.~Craps,
  E.~Keski-Vakkuri, B.~Muller, A.~Schafer, M.~Shigemori and W.~Staessens, {\it
  {Thermalization of Strongly Coupled Field Theories}},  {\em Phys. Rev. Lett.}
  {\bf 106} (2011) 191601 [\href{http://arXiv.org/abs/1012.4753}{{\tt
  1012.4753}}].

\bibitem{HaMa}
T.~Hartman and J.~Maldacena, {\it {Time Evolution of Entanglement Entropy from
  Black Hole Interiors}},  {\em JHEP} {\bf 05} (2013) 014
  [\href{http://arXiv.org/abs/1303.1080}{{\tt 1303.1080}}].

\bibitem{NNT}
M.~Nozaki, T.~Numasawa and T.~Takayanagi, {\it {Quantum Entanglement of Local
  Operators in Conformal Field Theories}},  {\em Phys. Rev. Lett.} {\bf 112}
  (2014) 111602 [\href{http://arXiv.org/abs/1401.0539}{{\tt 1401.0539}}].

\bibitem{Nozaki2014}
M.~Nozaki, {\it {Notes on Quantum Entanglement of Local Operators}},  {\em
  JHEP} {\bf 10} (2014) 147 [\href{http://arXiv.org/abs/1405.5875}{{\tt
  1405.5875}}].

\bibitem{Nozaki:2015mca}
M.~Nozaki, T.~Numasawa and S.~Matsuura, {\it {Quantum Entanglement of Fermionic
  Local Operators}},  {\em JHEP} {\bf 02} (2016) 150
  [\href{http://arXiv.org/abs/1507.04352}{{\tt 1507.04352}}].

\bibitem{Caputa2016b}
P.~Caputa, M.~Nozaki and T.~Numasawa, {\it {Charged Entanglement Entropy of
  Local Operators}},  {\em Phys. Rev.} {\bf D93} (2016), no.~10 105032
  [\href{http://arXiv.org/abs/1512.08132}{{\tt 1512.08132}}].

\bibitem{Nozaki:2016mcy}
M.~Nozaki and N.~Watamura, {\it {Quantum Entanglement of Locally Excited States
  in Maxwell Theory}},  {\em JHEP} {\bf 12} (2016) 069
  [\href{http://arXiv.org/abs/1606.07076}{{\tt 1606.07076}}].

\bibitem{HNTW}
S.~He, T.~Numasawa, T.~Takayanagi and K.~Watanabe, {\it {Quantum dimension as
  entanglement entropy in two dimensional conformal field theories}},  {\em
  Phys. Rev.} {\bf D90} (2014), no.~4 041701
  [\href{http://arXiv.org/abs/1403.0702}{{\tt 1403.0702}}].

\bibitem{CV}
P.~Caputa and A.~Veliz-Osorio, {\it {Entanglement constant for conformal
  families}},  {\em Phys. Rev.} {\bf D92} (2015), no.~6 065010
  [\href{http://arXiv.org/abs/1507.00582}{{\tt 1507.00582}}].

\bibitem{CGHW}
B.~Chen, W.-Z. Guo, S.~He and J.-q. Wu, {\it {Entanglement Entropy for
  Descendent Local Operators in 2D CFTs}},  {\em JHEP} {\bf 10} (2015) 173
  [\href{http://arXiv.org/abs/1507.01157}{{\tt 1507.01157}}].

\bibitem{Caputa2016yzn}
P.~Caputa and M.~M. Rams, {\it {Quantum dimensions from local operator
  excitations in the Ising model}},
  \href{http://arXiv.org/abs/1609.02428}{{\tt 1609.02428}}.

\bibitem{Numasawa2016}
T.~Numasawa, {\it {Scattering effect on entanglement propagation in RCFTs}},
  \href{http://arXiv.org/abs/1610.06181}{{\tt 1610.06181}}.

\bibitem{NNTH}
M.~Nozaki, T.~Numasawa and T.~Takayanagi, {\it {Holographic Local Quenches and
  Entanglement Density}},  {\em JHEP} {\bf 05} (2013) 080
  [\href{http://arXiv.org/abs/1302.5703}{{\tt 1302.5703}}].

\bibitem{CNT}
P.~Caputa, M.~Nozaki and T.~Takayanagi, {\it {Entanglement of local operators
  in large-N conformal field theories}},  {\em PTEP} {\bf 2014} (2014) 093B06
  [\href{http://arXiv.org/abs/1405.5946}{{\tt 1405.5946}}].

\bibitem{ABGH}
C.~T. Asplund, A.~Bernamonti, F.~Galli and T.~Hartman, {\it {Holographic
  Entanglement Entropy from 2d CFT: Heavy States and Local Quenches}},  {\em
  JHEP} {\bf 02} (2015) 171 [\href{http://arXiv.org/abs/1410.1392}{{\tt
  1410.1392}}].

\bibitem{Caputa2015b}
P.~Caputa, J.~Simon, A.~Stikonas, T.~Takayanagi and K.~Watanabe, {\it
  {Scrambling time from local perturbations of the eternal BTZ black hole}},
  {\em JHEP} {\bf 08} (2015) 011 [\href{http://arXiv.org/abs/1503.08161}{{\tt
  1503.08161}}].

\bibitem{Caputa2015c}
P.~Caputa, J.~Simon, A.~Stikonas and T.~Takayanagi, {\it {Quantum Entanglement
  of Localized Excited States at Finite Temperature}},  {\em JHEP} {\bf 01}
  (2015) 102 [\href{http://arXiv.org/abs/1410.2287}{{\tt 1410.2287}}].

\bibitem{CCT}
P.~Calabrese, J.~Cardy and E.~Tonni, {\it {Entanglement entropy of two disjoint
  intervals in conformal field theory}},  {\em J. Stat. Mech.} {\bf 0911}
  (2009) P11001 [\href{http://arXiv.org/abs/0905.2069}{{\tt 0905.2069}}].

\bibitem{Balasubramanian:2016xho}
V.~Balasubramanian, A.~Bernamonti, B.~Craps, T.~De~Jonckheere and F.~Galli,
  {\it {Entwinement in discretely gauged theories}},  {\em JHEP} {\bf 12}
  (2016) 094 [\href{http://arXiv.org/abs/1609.03991}{{\tt 1609.03991}}].

\bibitem{Shiba:2017vsr}
N.~Shiba, {\it {The Aharonov-Bohm Effect on Entanglement Entropy in Conformal
  Field Theory}},  \href{http://arXiv.org/abs/1701.00688}{{\tt 1701.00688}}.

\bibitem{Calabrese2013}
P.~Calabrese, J.~Cardy and E.~Tonni, {\it {Entanglement negativity in extended
  systems: A field theoretical approach}},  {\em J. Stat. Mech.} {\bf 1302}
  (2013) P02008 [\href{http://arXiv.org/abs/1210.5359}{{\tt 1210.5359}}].

\bibitem{Coser2014}
A.~Coser, L.~Tagliacozzo and E.~Tonni, {\it {On R\'enyi entropies of disjoint
  intervals in conformal field theory}},  {\em J. Stat. Mech.} {\bf 1401}
  (2014) P01008 [\href{http://arXiv.org/abs/1309.2189}{{\tt 1309.2189}}].

\bibitem{Headrick:2010zt}
M.~Headrick, {\it {Entanglement Renyi entropies in holographic theories}},
  {\em Phys. Rev.} {\bf D82} (2010) 126010
  [\href{http://arXiv.org/abs/1006.0047}{{\tt 1006.0047}}].

\bibitem{Hartman:2013mia}
T.~Hartman, {\it {Entanglement Entropy at Large Central Charge}},
  \href{http://arXiv.org/abs/1303.6955}{{\tt 1303.6955}}.

\bibitem{Asplund2015a}
C.~T. Asplund, A.~Bernamonti, F.~Galli and T.~Hartman, {\it {Entanglement
  Scrambling in 2d Conformal Field Theory}},  {\em JHEP} {\bf 09} (2015) 110
  [\href{http://arXiv.org/abs/1506.03772}{{\tt 1506.03772}}].

\bibitem{Blanco:2011np}
D.~D. Blanco and H.~Casini, {\it {Entanglement entropy for non-coplanar regions
  in quantum field theory}},  {\em Class. Quant. Grav.} {\bf 28} (2011) 215015
  [\href{http://arXiv.org/abs/1103.4400}{{\tt 1103.4400}}].

\bibitem{Roberts2015}
D.~A. Roberts and D.~Stanford, {\it {Two-dimensional conformal field theory and
  the butterfly effect}},  {\em Phys. Rev. Lett.} {\bf 115} (2015), no.~13
  131603 [\href{http://arXiv.org/abs/1412.5123}{{\tt 1412.5123}}].

\bibitem{Kitaev}
A.~Kitaev, ``Hidden correlation in the hawking radiation and thermal noise.''
  Talk given at the Fundamental Physics Prize Symposium, Nov.10, 2014.

\bibitem{Maldacena2016}
J.~Maldacena, S.~H. Shenker and D.~Stanford, {\it {A bound on chaos}},  {\em
  JHEP} {\bf 08} (2016) 106 [\href{http://arXiv.org/abs/1503.01409}{{\tt
  1503.01409}}].

\bibitem{CKTW}
P.~Caputa, Y.~Kusuki, T.~Takayanagi and K.~Watanabe, {\it {Out-of-Time-Ordered
  Correlators in $(T^2)^n/\mathbb{Z}_n$}},
  \href{http://arXiv.org/abs/1703.09939}{{\tt 1703.09939}}.

\bibitem{CNV}
P.~Caputa, T.~Numasawa and A.~Veliz-Osorio, {\it {Out-of-time-ordered
  correlators and purity in rational conformal field theories}},  {\em PTEP}
  {\bf 2016} (2016), no.~11 113B06 [\href{http://arXiv.org/abs/1602.06542}{{\tt
  1602.06542}}].

\bibitem{Gu2016}
Y.~Gu and X.-L. Qi, {\it {Fractional Statistics and the Butterfly Effect}},
  {\em JHEP} {\bf 08} (2016) 129 [\href{http://arXiv.org/abs/1602.06543}{{\tt
  1602.06543}}].

\bibitem{Bantay1998}
P.~Bantay, {\it {Characters and modular properties of permutation orbifolds}},
  {\em Phys. Lett.} {\bf B419} (1998) 175--178
  [\href{http://arXiv.org/abs/hep-th/9708120}{{\tt hep-th/9708120}}].

\bibitem{Drinfeld1986}
V.~G. Drinfeld, {\it {Quantum groups}},  {\em J. Sov. Math.} {\bf 41} (1988)
  898--915. [Zap. Nauchn. Semin.155,18(1986)].

\bibitem{JevickiYoon2016}
A.~Jevicki and J.~Yoon, {\it {$S_N$ Orbifolds and String Interactions}},  {\em
  J. Phys.} {\bf A49} (2016), no.~20 205401
  [\href{http://arXiv.org/abs/1511.07878}{{\tt 1511.07878}}].

\end{thebibliography}\endgroup

\end{document}